\documentclass[pra,aps,showpacs,11pt,notitlepage]{revtex4-1}

\usepackage{amsmath,amsfonts,amssymb,caption,color,epsfig,graphics,graphicx,latexsym,mathrsfs,revsymb,theorem,url,verbatim,epstopdf}

\usepackage{lineno,hyperref}
\newtheorem{definition}{Definition}
\newtheorem{proposition}[definition]{Proposition}
\newtheorem{lemma}[definition]{Lemma}

\newtheorem{theorem}[definition]{Theorem}
\newtheorem{corollary}[definition]{Corollary}
\newtheorem{conjecture}[definition]{Conjecture}

\newtheorem{remark}[definition]{Remark}
\newtheorem{example}[definition]{Example}
\newtheorem{question}[definition]{Question}

\def\squareforqed{\hbox{\rlap{$\sqcap$}$\sqcup$}}
\def\qed{\ifmmode\squareforqed\else{\unskip\nobreak\hfil
\penalty50\hskip1em\null\nobreak\hfil\squareforqed
\parfillskip=0pt\finalhyphendemerits=0\endgraf}\fi}
\def\endenv{\ifmmode\;\else{\unskip\nobreak\hfil
\penalty50\hskip1em\null\nobreak\hfil\;
\parfillskip=0pt\finalhyphendemerits=0\endgraf}\fi}
\newenvironment{proof}{\noindent \textbf{{Proof.~} }}{\qed}
\def\Dbar{\leavevmode\lower.6ex\hbox to 0pt
{\hskip-.23ex\accent"16\hss}D}
\makeatletter
\def\url@leostyle{%
  \@ifundefined{selectfont}{\def\UrlFont{\sf}}{\def\UrlFont{\small\ttfamily}}}
\makeatother
\def\bcj{\begin{conjecture}}
\def\ecj{\end{conjecture}}
\def\bcr{\begin{corollary}}
\def\ecr{\end{corollary}}
\def\bd{\begin{definition}}
\def\ed{\end{definition}}
\def\bea{\begin{eqnarray}}
\def\eea{\end{eqnarray}}
\def\bem{\begin{enumerate}}
\def\eem{\end{enumerate}}
\def\bex{\begin{example}}
\def\eex{\end{example}}
\def\bim{\begin{itemize}}
\def\eim{\end{itemize}}
\def\bl{\begin{lemma}}
\def\el{\end{lemma}}
\def\bpf{\begin{proof}}
\def\epf{\end{proof}}
\def\bpp{\begin{proposition}}
\def\epp{\end{proposition}}
\def\bqu{\begin{question}}
\def\equ{\end{question}}
\def\br{\begin{remark}}
\def\er{\end{remark}}
\def\bt{\begin{theorem}}
\def\et{\end{theorem}}

\def\btb{\begin{tabular}}
\def\etb{\end{tabular}}

\newcommand{\nc}{\newcommand}

\def\a{\alpha}
\def\b{\beta}

\def\d{\delta}

\def\t{\theta}

\def\p{\pi}

\def\s{\sigma}

\def\ph{\varphi}

\def\Ps{\Psi}

 \nc{\bA}{{\bf A}} \nc{\bB}{{\bf B}} \nc{\bC}{{\bf C}}
 \nc{\bD}{{\bf D}} \nc{\bE}{{\bf E}} \nc{\bF}{{\bf F}}
 \nc{\bG}{{\bf G}} \nc{\bH}{{\bf H}} \nc{\bI}{{\bf I}}
 \nc{\bJ}{{\bf J}} \nc{\bK}{{\bf K}} \nc{\bL}{{\bf L}}
 \nc{\bM}{{\bf M}} \nc{\bN}{{\bf N}} \nc{\bO}{{\bf O}}
 \nc{\bP}{{\bf P}} \nc{\bQ}{{\bf Q}} \nc{\bR}{{\bf R}}
 \nc{\bS}{{\bf S}} \nc{\bT}{{\bf T}} \nc{\bU}{{\bf U}}
 \nc{\bV}{{\bf V}} \nc{\bW}{{\bf W}} \nc{\bX}{{\bf X}}
 \nc{\bZ}{{\bf Z}}

\nc{\cA}{{\cal A}} \nc{\cB}{{\cal B}} \nc{\cC}{{\cal C}}
\nc{\cD}{{\cal D}} \nc{\cE}{{\cal E}} \nc{\cF}{{\cal F}}
\nc{\cG}{{\cal G}} \nc{\cH}{{\cal H}} \nc{\cI}{{\cal I}}
\nc{\cJ}{{\cal J}} \nc{\cK}{{\cal K}} \nc{\cL}{{\cal L}}
\nc{\cM}{{\cal M}} \nc{\cN}{{\cal N}} \nc{\cO}{{\cal O}}
\nc{\cP}{{\cal P}} \nc{\cQ}{{\cal Q}} \nc{\cR}{{\cal R}}
\nc{\cS}{{\cal S}} \nc{\cT}{{\cal T}} \nc{\cU}{{\cal U}}
\nc{\cV}{{\cal V}} \nc{\cW}{{\cal W}} \nc{\cX}{{\cal X}}
\nc{\cZ}{{\cal Z}}

\nc{\hA}{{\hat{A}}} \nc{\hB}{{\hat{B}}} \nc{\hC}{{\hat{C}}}
\nc{\hD}{{\hat{D}}} \nc{\hE}{{\hat{E}}} \nc{\hF}{{\hat{F}}}
\nc{\hG}{{\hat{G}}} \nc{\hH}{{\hat{H}}} \nc{\hI}{{\hat{I}}}
\nc{\hJ}{{\hat{J}}} \nc{\hK}{{\hat{K}}} \nc{\hL}{{\hat{L}}}
\nc{\hM}{{\hat{M}}} \nc{\hN}{{\hat{N}}} \nc{\hO}{{\hat{O}}}
\nc{\hP}{{\hat{P}}} \nc{\hR}{{\hat{R}}} \nc{\hS}{{\hat{S}}}
\nc{\hT}{{\hat{T}}} \nc{\hU}{{\hat{U}}} \nc{\hV}{{\hat{V}}}
\nc{\hW}{{\hat{W}}} \nc{\hX}{{\hat{X}}} \nc{\hZ}{{\hat{Z}}}

\nc{\hn}{{\hat{n}}}

\def\diag{\mathop{\rm diag}}
\def\dim{\mathop{\rm Dim}}

\def\lin{\mathop{\rm span}}

\def\max{\mathop{\rm max}}
\def\min{\mathop{\rm min}}

\def\rank{\mathop{\rm rank}}

\def\tr{\mathop{\rm Tr}}

\def\dg{\dagger}

\def\lc{\lceil}
\def\rc{\rceil}

\def\op{\oplus}
\def\ox{\otimes}

\def\ra{\rightarrow}

\def\su{\subset}
\def\sue{\subseteq}
\def\sm{\setminus}

\newcommand{\bra}[1]{\langle#1|}
\newcommand{\ket}[1]{|#1\rangle}
\newcommand{\proj}[1]{| #1\rangle\!\langle #1 |}
\newcommand{\ketbra}[2]{|#1\rangle\!\langle#2|}
\newcommand{\braket}[2]{\langle#1|#2\rangle}

\newcommand{\norm}[1]{\lVert#1\rVert}
\newcommand{\abs}[1]{|#1|}

\def\Dbar{\leavevmode\lower.6ex\hbox to 0pt
{\hskip-.23ex\accent"16\hss}D}

\begin{document}

\title{On the Schmidt-rank-three bipartite and multipartite unitary operator}





\author{ Lin Chen $^{1}$ and Li Yu $^{1}$\\
\small ${}^{1}$ Singapore University of Technology and Design, 20
Dover Drive, Singapore 138682
}

\begin{abstract}
Unitary operations are physically implementable.  We further the understanding of such operations by studying the possible forms of nonlocal unitary operators, which are bipartite or multipartite unitary operators that are not tensor product operators. They are of broad relevance in quantum information processing. We prove that any nonlocal unitary operator of Schmidt rank three on a $d_A\times d_B$ bipartite system is locally equivalent to a controlled unitary. This operator can be locally implemented assisted by a maximally entangled state of Schmidt rank $\min\{d_A^2,d_B\}$ when $d_A\le d_B$. We further show that any multipartite unitary operator $U$ of Schmidt rank three can be controlled by one system or collectively controlled by two systems, regardless of the number of systems of $U$. In the scenario of $n$-qubit, we construct non-controlled $U$ for any odd $n\ge5$, and prove that $U$ is a controlled unitary for any even $n\ge4$.
\end{abstract}






\maketitle

\section{Introduction}
\label{sec:intro}

Bipartite and multipartite unitary operators play a fundamental role in quantum information processing. They are used to create quantum states, nonlocal correlations such as entanglement and discord \cite{oz01}, and to implement quantum circuits and computation. It is thus desirable to systematically implement multipartite unitary operators and understand their properties. So far this is a hard problem.
The understanding of the forms and implementation schemes of multipartite unitary operators is still far from complete.
For a survey of the literature on this topic, we refer the readers to the introductory parts of \cite{cy13} and \cite{cy14}, but in the next paragraph we review some basic concepts and known facts about the nonlocal unitary operators and the controlled unitary operators.

Controlled unitary operators are a subclass of multipartite unitary operators. For example, the product of the local unitary operators on the local systems is the simplest controlled unitary. Such a unitary operator with \textit{Schmidt rank} one is a local operator, also known as a product operator. Otherwise it is a \textit{nonlocal} operator. Nonlocal unitaries can create quantum entanglement between distributed parties \cite{ejp00}, and their equivalence has been studied under local operations and classical communication (LOCC) \cite{dc02}. Without prior entanglement, nonlocal unitaries cannot be implemented by LOCC only, even if the probability is allowed to be close to zero \cite{pv98}. To implement these tasks, it is desirable to have a simple type of nonlocal unitaries. The bipartite controlled unitary gates turn out to be such a type. They are of the general form $U=\sum_{j=1}^m P_j \ox V_j$ acting on a bipartite Hilbert space $\cH_A \ox \cH_B$, where $P_j$ are orthogonal projectors on $\cH_A$ and $V_j$ are unitaries on $\cH_B$. They can be implemented by a simple nonlocal protocol \cite{ygc10} using a maximally entangled state of Schmidt rank $m$. In this sense the implementation of controlled unitaries is operational. The bipartite controlled unitary gates are one of the few classes of bipartite unitaries for which their capacity to create entanglement between the parts is relatively well understood \cite{WangSandersBerry03}, and \cite{stm11,sg11} contain significant progress toward understanding their entanglement cost. Controlled unitaries also play an important role in quantum information theory, e.g. they are used in gate sets that are universal \cite{bbc95} for quantum computation \cite{nc2000book}, and they were also used in the creation of graph states (cluster states) \cite{Sch03}, which find wide use in quantum communication protocols \cite{BnBr84} and quantum computation.

\medskip

\begin{figure}[ht]
\center{\includegraphics{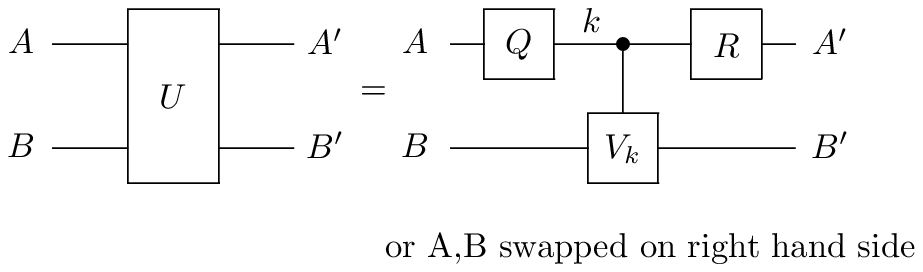}}
\captionsetup{font=footnotesize,width=0.95\textwidth}
\caption{Any bipartite unitary $U$ of Schmidt rank three is locally equivalent to a controlled unitary, where the controlling side may be $A$ or $B$.  This is expressed as $U=(Q\ox I)(\sum^{d_A}_{k=1}\proj{k}\ox V_k)(R\ox I)$ or $U=(I\ox Q)(\sum^{d_B}_{k=1}V_k \ox \proj{k})(I\ox R)$, where $V_k$, $Q$ and $R$ are local unitaries. The output systems $A'$ and $B'$ are assumed to be of the same size as $A$ and $B$, respectively.}
\label{fig1}
\end{figure}

\begin{figure}[ht]
\center{\includegraphics{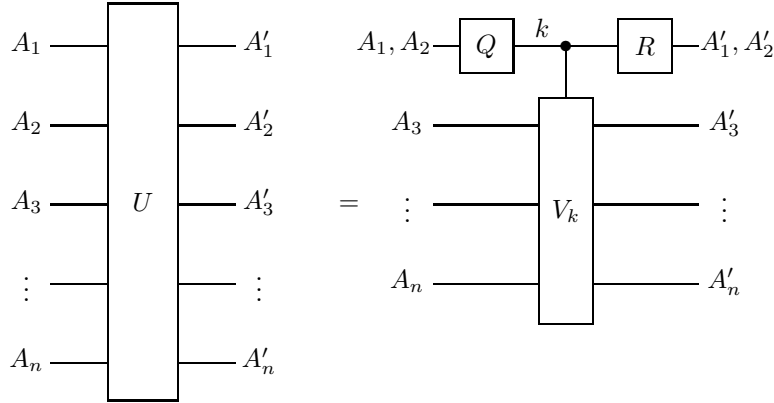}}
\captionsetup{font=footnotesize,width=0.95\textwidth}
\caption{Any $n$-partite unitary $U$ of Schmidt rank three is locally equivalent to a controlled unitary by one system, or a unitary collectively controlled by two systems (but the systems are not arbitrary). The case that $U$ is controlled by two systems is illustrated in the figure, and the case of one system acting as control is similar. The two controlling systems are denoted as $A_1,A_2$.  And $U=(Q\ox I)(\sum^{d}_{k=1}\proj{k}\ox V_k)(R\ox I)$, where $Q$ and $R$ are unitary operations on the space $\cH_1\ox\cH_2$ of dimension $d$, and $V_k$ are unitary operations on $\cH_3\ox\cdots\ox\cH_n$. The output systems $A_1',\cdots,A_n'$ are assumed to be of the same size as $A_1,\cdots,A_n$, respectively.}
\label{fig2}
\end{figure}

In this paper, we study this problem via the Schmidt-rank-three bipartite and multipartite unitary operators. We show that any Schmidt-rank-three bipartite unitary operator on $d_A \times d_B$ system is locally
equivalent to a controlled unitary operator, see Theorem~\ref{thm:sch3} and Fig.~\ref{fig1}. The theorem is proved by developing novel tools from linear algebra in Sec.~\ref{subsec:linearalg}.
Our results directly generalize most findings and applications in \cite{cy14}. In particular, we show that the unitary can be implemented by
LOCC and a maximally entangled state $\ket{\Ps_r}={1\over \sqrt
r}\sum^r_{i=1}\ket{ii}$, where $r=\min\{d_A^2,d_B\}$ and $d_A\le d_B$, see Lemma \ref{le:costSR3}.
As an application different from those in \cite{cy14}, we show that any multipartite unitary operator of Schmidt rank three can be controlled by one system or collectively controlled by two systems, regardless of the number of systems of this operator. This is illustrated in Fig. \ref{fig2} and Theorem \ref{thm:multi}. The theorem is further strengthened in Corollary \ref{cr:multi}, by which we show that every Schmidt-rank-three multipartite unitary is controlled by the union of two systems (such two systems are considered as one combined system). By ``collectively controlled by $m$ systems'' we mean that, the appointed $m$ systems are regarded as one system which controls the multipartite unitary, and the unitary is not controlled by any $k$ ($k\le m-1$) systems.

We also study the Schmidt-rank-three $n$-qubit unitary $U$. For any odd $n\ge3$, we present examples of $U$ that are collectively controlled by two systems, see Eqs.~\eqref{eq:sr3threequbit} and \eqref{eq:sr3nqubit}. For any even $n\ge4$, $U$ turns out to be a controlled unitary, see Proposition~\ref{pp:evenqubit}. We also introduce a connection between the controlled unitary and Schmidt rank in Lemma~\ref{le:schnum}. The algebra tools we develop also reproduce the main result in \cite{cy13}, see Lemma~\ref{le:sch2}.

The rest of this paper is organized as follows. In Sec.~\ref{sec:pre} we introduce the preliminary knowledge. In
Sec.~\ref{sec:sch3} we prove our main result that any Schmidt-rank-three bipartite unitary is a controlled unitary. Next we propose a few applications of the main result in Sec.~\ref{sec:app}. We show that any Schmidt-rank-three multipartite unitary is either a controlled unitary, or collectively controlled by two systems. In particular, any Schmidt-rank-three $n$-qubit unitary with any even $n\ge4$ is a controlled unitary. We also discuss the implication of our main result to the entanglement cost of bipartite unitaries, the connection between the Schmidt rank of a unitary and the Schmidt number of some related quantum states, and give a simple proof for that any Schmidt-rank-two bipartite unitary is a controlled unitary.
Finally, we conclude in Sec.~\ref{sec:conclusion}.

\section{Preliminaries}\label{sec:pre}

Let $\cH=\cH_A\otimes\cH_B$ be the complex Hilbert space of a
finite-dimensional bipartite quantum system of Alice and Bob. We
denote by $d_A,d_B$ the dimension of $\cH_A$ and $\cH_B$,
respectively. It is known that $\cH$ is spanned by the computational
basis $\ket{i,j}$, where $i=1,\cdots,d_A$, and $j=1,\cdots,d_B$. We shall denote
$I_k=\sum^k_{i=1} \proj{i}$. For convenience, we denote
$I_A=I_{d_A}$, $I_B=I_{d_B}$ and $I=I_{d_Ad_B}$ as the identity
operator on spaces $\cH_A,\cH_B$, and $\cH$, respectively. Two bipartite unitary operators $U,V$ on $\cH$ are \textit{locally equivalent} when there are two product unitaries $S_1,S_2$ such that
$U=S_1VS_2$.
We say that $U$ is a \textit{controlled unitary gate}, if $U$ is
locally equivalent to $\sum^{d_A}_{j=1}\proj{j}\ox U_j$ or
$\sum^{d_B}_{j=1}V_j \ox \proj{j}$. To be specific, $U$ is a
controlled unitary from $A$ or $B$ side.
Next, $U$ has Schmidt rank $n$ if there is a Schmidt decomposition (or expansion) $U=\sum^n_{j=1}A_j \ox B_j$ where the $d_A\times d_A$ matrices $A_1,\cdots,A_n$ are linearly independent, and the
$d_B\times d_B$ matrices $B_1,\cdots,B_n$ are also linearly independent. In particular, the zero matrix has Schmidt rank zero.
We name the $A$ $(B)$ space of $U$ as the space spanned by all $A_j$ $(B_j)$ that appear in a Schmidt decomposition of $U$. It is well defined in the sense that the space is independent of the specific choice of the Schmidt decomposition. For example, the identity operator $I$ is the simplest controlled unitary and has Schmidt rank one.

Let us recall the concept of block-controlled unitary gate \cite{cy14}. We split the space into a direct sum:
$\cH_A=\op^m_{i=1} \cH_i$, $m>1$, $\dim\cH_i=m_i$, and
$\cH_i\perp\cH_j$ for distinct $i,j=1,\cdots,m$. We say that $U$ is
a \textit{block-controlled unitary (BCU) gate controlled from the A
side}, if $U$ is locally equivalent to $\sum^m_{i=1}
\sum^{m_i}_{j,k=1} \ketbra{u_{ij}}{u_{ik}}\ox V_{ijk}$ where
$\ket{u_{i,1}},\cdots,\ket{u_{i,m_i}}$ is an orthonormal basis of $\cH_i$. Note that the $V_{ijk}$ are not necessarily unitary. For simplicity we
denote the decomposition as $\op_A V_i$ where
$V_i=\sum^{m_i}_{j,k=1} \ketbra{u_{ij}}{u_{ik}}\ox U_{ijk}$, and we
denote $\abs{V_i}_A=m_i$. We have $UU^\dg=\sum^m_{i=1} P_{\cH_i} \ox
I_{B}=I$. In this paper we give an alternative and more easily accessible definition of BCU as follows. A unitary $U$ is a BCU controlled from the $A$ side if and only if $U=\sum_i \ketbra{a_i}{b_i} \ox V_i + \sum_j \ketbra{c_j}{e_j} \ox W_j$ where $\braket{a_i}{c_j}=\braket{b_i}{e_j}=0$, $\forall i,j$. One may
similarly define the BCU gate controlled from the $B$ side. We shall use both definitions in the remaining part of this paper, and say $U$ is a BCU when it is from either $A$ or $B$ side. By definition every controlled unitary with the condition $d_Ad_B>1$ is a BCU. As this condition is generically satisfied, we will simply state that every controlled unitary is a BCU.

Not every bipartite unitary is a BCU. For example, the SWAP gate $\frac{1}{2}\sum^3_{i=0}\s_i\ox\s_i$ acts on two-qubit space and has Schmidt rank 4. Here $\s_0=I_2$, and
\bea
\s_1=\left(
                   \begin{array}{cc}
                     0 & 1 \\
                     1 & 0 \\
                     \end{array}
                 \right),
~~
\s_2=\left(
                   \begin{array}{cc}
                     0 & -i \\
                     i & 0 \\
                     \end{array}
                 \right),
~~
\s_3=\left(
                   \begin{array}{cc}
                     1 & 0 \\
                     0 & -1 \\
                     \end{array}
                 \right),
\eea
are standard Pauli matrices. By definition, the SWAP gate is neither a controlled unitary nor a BCU. We will refer to $W\op_A X$ as \textit{the direct sum of matrices $W$ and $X$ from the A side}, or A-direct sum of $W$ and $X$. Here $W$, $X$ respectively act on the subspaces $H_i\ox\cH_B$, $i=1,2$, where $H_1\perp H_2$. Indeed, both controlled unitaries and BCU are the direct sum of some blocks.

\subsection{Preliminary results on linear algebra}
\label{subsec:linearalg}

We present a few preliminary lemmas based on linear algebra. They are useful for our main result Theorem \ref{thm:sch3}, as well as the investigation of multipartite unitaries in Sec. \ref{sec:app}.

\bl\label{le:normal}
Let $A$ be a square matrix and $AA^\dg=\sum^k_{i=1} c_i P_i$, where the $c_i$ are distinct nonnegative real numbers, and the projectors $P_i$ are pairwise orthogonal. Then
\\
(i) $A=(\sum^k_{i=1} c_i^{\frac12} P_i)V$, where $V$ is a unitary matrix.
\\
(ii) If furthermore $A$ is normal, then $A= \op^k_{i=1} c_i^{\frac12} U_i$, where the $U_i$ is a unitary on the subspace of $P_i$.
\\
(iii) Let $W,X$ be two unitary matrices and $W(\sum^k_{i=1} c_i P_i)X=\sum^k_{i=1} c_i P_i$. Then $W=\op^k_{i=1} W_i$ and $X=\op^k_{i=1} X_i$, where the $W_i$ and $X_i$ are two unitary matrices on the subspace of $P_i$, and $W_i=X_i^\dg$ for all $i$ such that $c_i$ is nonzero.
\\
(iv) Let $W,X$ be two unitary matrices and $WDX=D$, where $D$ is a diagonal positive definite matrix. Then $W=X^\dg$.
\el
\bpf
(i) The assertion is known in matrix theory.

(ii) Since $A$ is normal and $AA^\dg=\sum^k_{i=1} c_i P_i$, we have $A=U(\sum^k_{i=1} {c_i}^{\frac12} P_i')U^\dg$ where $U$ is a unitary and $P_i'$ is a diagonal unitary on the space of $P_i$. Hence $P_i' (P_j')^\dg = \d_{ij} P_i = P_iP_j$. It follows from $AA^\dg=\sum^k_{i=1} c_i P_i$ that $U$ and $\sum^k_{i=1} c_i P_i$ commute. Since the $c_i$ are pairwise different, we have $U=V_1\op \cdots \op V_k$, where the $V_i$ is a unitary on the space of $P_i$. Letting $U_i=V_iP_i'V_i^\dg$ for all $i$ leads to the assertion.

(iii) Since $W(\sum^k_{i=1} c_i P_i)X=\sum^k_{i=1} c_i P_i$, we have $X^\dg(\sum^k_{i=1} c_i P_i)W^\dg=\sum^k_{i=1} c_i P_i$. Their products are
\bea
\label{eq:iii}
W\bigg(\sum^k_{i=1} c_i^2 P_i \bigg)W^\dg
=
X^\dg\bigg(\sum^k_{i=1} c_i^2 P_i \bigg)X
= \sum^k_{i=1} c_i^2 P_i.
\eea
Since the $c_i$ are nonnegative, real and pairwise different, so are the $c_i^2$. Eq. \eqref{eq:iii} implies that $W=\op^k_{i=1} W_i$ and $X=\op^k_{i=1} X_i$, where the $W_i$ and $X_i$ are two unitary matrices on the subspace of $P_i$. Since $W(\sum^k_{i=1} c_i P_i)X=\sum^k_{i=1} c_i P_i$, we have $c_iW_iX_i=c_iP_i$ for any $i$. Thus $W_i=X_i^\dg$ for all $i$ such that $c_i$ is nonzero. So assertion (iii) follows.

(iv) Since $D$ is a diagonal positive definite matrix, there is a permutation matrix $P$ such that $D=P(\sum^k_{i=1} c_i P_i)P^\dg$.
Since $WDX=D$, we have
\bea
P^\dg W P \bigg( \sum^k_{i=1} c_i P_i \bigg)P^\dg X P
=
\sum^k_{i=1} c_i P_i.
\eea
Since $D$ is positive definite, (iii) implies that $P^\dg W P=(P^\dg X P)^\dg$. Hence $W=X^\dg$. This completes the proof.
\epf

\bl\label{le:singular}
Any space spanned by two linearly independent square matrices contains a nonzero singular matrix.
\el
\bpf
Denote the two matrices as $A$ and $B$. If the assertion does not hold, then $A$ and $B$ are both invertible. Let $\lambda$ be an eigenvalue of $A^{-1} B$, we have $\det(-\lambda I_A+A^{-1}B)=0$, hence $\det(-\lambda A+B)=0$, meaning that $C:=-\lambda A + B$ is singular. Since $A$ and $B$ are linearly independent, $C$ is nonzero. This completes the proof.
\epf

\bl
\label{le:orthogonalization}
Let $A_1,A_2$ be two linearly independent square matrices. Let $x_i,y_i$ be two positive numbers, $z_i$ one complex number, such that $x_iy_i>\abs{z_i}^2$ for $i=1,2$ and
\bea
\label{eq:orthogonalization1}
x_1A_1^\dg A_1 + y_1A_2^\dg A_2 + z_1 A_1^\dg A_2 + z_1^* A_2^\dg A_1 = I,
\notag\\
\label{eq:orthogonalization2}
x_2A_1 A_1^\dg + y_2A_2 A_2^\dg + z_2 A_1 A_2^\dg + z_2^* A_2 A_1^\dg = I.
\eea
Then the following statements hold:

(i) There are two positive numbers $a,b$, and two linearly independent matrices $B_1,B_2$ in the span of $A_1,A_2$ such that
\bea
\label{eq:orthogonalization3}
B_1^\dg B_1 + B_2^\dg B_2 = I,
\notag\\
\label{eq:orthogonalization4}
a B_1 B_1^\dg + b B_2 B_2^\dg = I.
\eea
(ii) If one of $B_1$ and $B_2$ is not proportional to a unitary, then $a=b=1$.
\\
(iii) $B_1$ and $B_2$ always have simultaneous singular value decomposition, i.e., there are unitary matrices $V,W$ such that $VB_1W$ and $VB_2W$ both diagonal.
\el
\bpf
(i) Since $x_iy_i>\abs{z_i}^2$ for $i=1,2$, then \eqref{eq:orthogonalization1} implies that there are two linearly independent matrices $A_3=(x_1-{\abs{z_1}^2\over y_1})^{\frac12}A_1$, and $A_4$ in the space $H=\lin\{A_1,A_2\}$ such that
\bea
\label{eq:orthogonalization5}
&&A_3^\dg A_3 + A_4^\dg A_4=I,
\notag\\
\label{eq:orthogonalization6}
&&
x A_3 A_3^\dg + y A_4 A_4^\dg + z A_3 A_4^\dg + z^* A_4 A_3^\dg = I,
\eea
with two positive numbers $x,y$, one complex number $z=\abs{z}e^{-i\ph}$ and a real phase $\ph$. It follows from \eqref{eq:orthogonalization1} and $x_2y_2>\abs{z_2}^2$ that $xy>\abs{z}^2$. If $x=y$, then \eqref{eq:orthogonalization5} implies that $a=x+\abs{z}$, $b=x-\abs{z}$, and $B_{1,2}={1\over\sqrt2}A_3\pm{1\over\sqrt2}e^{i\ph} A_4$ in \eqref{eq:orthogonalization3}. Since $x>\abs{z}$, we have $a,b>0$. Since $B_1,B_2$ span $H$, the assertion follows. It suffices to assume $x\ne y$. Let
\bea
&& B_1 = \cos\t A_3 + e^{i\ph} \sin\t A_4,
\\
&& B_2 = \sin\t A_3 - e^{i\ph} \cos\t A_4,
\\
&& a= {1\over2} (x+y + \text{sgn}(x-y) \sqrt{(x-y)^2+ 4\abs{z}^2}),
\\
\label{eq:orthogonalization7}
&& b= {1\over2} (x+y - \text{sgn}(x-y) \sqrt{(x-y)^2+ 4\abs{z}^2}),
\eea
where $\t={1\over2}\arctan{2\abs{z} \over x-y}$, and $\text{sgn}()$ is the sign function. One can easily verify that these expressions make \eqref{eq:orthogonalization3} and \eqref{eq:orthogonalization5} the same. It follows from \eqref{eq:orthogonalization7} and $xy>\abs{z}^2$ that $a,b>0$. Since $B_1,B_2$ span $H$, the assertion follows.

(ii) Up to local unitaries on the l. h. s. of \eqref{eq:orthogonalization3}, we may assume that $B_1=\sum^k_{i=1} c_i^{\frac12} P_i$, where the $c_i$ are nonnegative, real and $c_i>c_{i+1}$ for all $i$, and the projectors $P_i$ are pairwise orthogonal. Eq. \eqref{eq:orthogonalization3} implies that $B_2^\dg B_2 =I - B_1^\dg B_1$ and $B_2 B_2^\dg ={1\over b} I - {a \over b} B_1 B_1 ^\dg $. The r.h.s. of both equations are diagonal matrices whose diagonals are in the ascending order. Since $B_2 B_2^\dg$ and $B_2^\dg B_2$ are similar matrices, they have identical eigenvalues. We have $B_2 B_2^\dg=B_2^\dg B_2=\sum^k_{i=1} (1-c_i)P_i$. Since one of $B_1$ and $B_2$ is not proportional to a unitary, we have $k>1$.
Then Eq. \eqref{eq:orthogonalization3} implies that $a=b=1$.

(iii) From (ii), we have that either $B_1,B_2$ are both proportional to unitaries, or none of $B_1,B_2$ are proportional to unitaries but they satisfy Eq.~\eqref{eq:orthogonalization3} with $a=b=1$.
In the former case, up to unitaries before and after the operators $B_1$ and $B_2$, we may assume $B_1\propto I_B$, and $B_2$ is diagonal, hence $B_1$ and $B_2$ have simultaneous singular value decomposition. In the latter case, up to unitaries we may assume $B_1$ is a diagonal matrix, i.e., $B_1=\sum^k_{i=1} c_i^{\frac12}P_i$, where the $c_i$ are nonnegative, real, pairwise different, and the projectors $P_i$ are pairwise orthogonal. From Eq.~\eqref{eq:orthogonalization3} with $a=b=1$, we have $B_2 B_2^\dg=B_2^\dg B_2=\sum^k_{i=1} (1-c_i)P_i$. So $B_2$ is normal. It follows from Lemma \ref{le:normal} that $B_2= \op^k_{i=1} (1-c_i)^{\frac12} U_i$, where the $U_i$ is a unitary on the subspace of $P_i$, and can be diagonalized simultaneously with $P_i$. Hence in the latter case, $B_1$ and $B_2$ also have simultaneous singular value decomposition. This completes the proof.
\epf

\bl
\label{le:schineq}
Suppose a bipartite operator $\sum_{j=1}^N A_j\ox B_j$ has Schmidt rank $r\ge 0$, and
\bea
\delta_A:=\dim\left({\rm span}\{A_j\}_{j=1}^N\right),
\\
\delta_B:=\dim\left({\rm span}\{B_j\}_{j=1}^N\right).
\eea Then
\\
(i) $\delta_A+\delta_B\le N+r$;
\\
(ii) $r\le \min\{\delta_A, \delta_B\}\le \max\{\delta_A, \delta_B\}\le N$;
\\
(iii) If $\max\{\delta_A, \delta_B\}=N$, then $\min\{\delta_A, \delta_B\}=r$.
\el
\bpf
(i) The assertion is from \cite{cohen12}.

(ii) It is sufficient to prove the first inequality. If $r> \min\{\delta_A, \delta_B\}$, then the Schmidt rank of the bipartite operator is smaller than $r$. It gives us a contradiction, so the assertion follows.

(iii) The assertion follows from the definition of Schmidt rank. This completes the proof.
\epf

The converse of (iii) is wrong. If $\min\{\delta_A, \delta_B\}=r$, then we have $\max\{\delta_A, \delta_B\}\le N$.
\smallskip

The equivalence between a bipartite and controlled unitary has been widely studied recently \cite{cy13,cy14}. For the purpose in this paper, we provide more equivalent conditions of deciding whether a bipartite unitary is a controlled unitary.
 \bl
 \label{le:controlunitary}
Let $U=\sum_j A_j \ox B_j$ be a bipartite unitary and consider the
following five assertions:
\\
(i) $U$ is a controlled unitary from the $A$ side;
\\
(ii) there are two orthonormal basis $\{\ket{a_i}\}$ and $\{\ket{b_i}\}$ of $\cH_A$ such that $U=\sum_{i}(\proj{a_i}\ox I_B)U(\proj{b_i}\ox I_B)$;
\\
(iii) for two arbitrary orthonormal basis $\{\ket{c_i}\}$ and $\{\ket{e_j}\}$ of $\cH_B$, there are unitaries $S$ and $T$ on $\cH_A$ such that the operators $(S\ox\bra{c_i})U(T\ox\ket{e_j})$ on $\cH_A$, $\forall i,j$ are all diagonal;
\\
(iv) the matrices $A_i$ have simultaneous singular value decomposition.
\\
(v) the operators $A_i A_j^\dg$, $\forall i,j$ are all normal and commute with
each other, and the operators $A_i^\dg A_j$, $\forall i, j$ are all normal
and commute with each other.

Then the first three assertions are equivalent. If the operators $\{B_j\}$ are linearly independent, then all five assertions are equivalent.
 \el
\bpf
The relation $(i)\ra(ii),(iii)$ follows from the definition of controlled unitaries. Let us prove the relation $(ii)\ra(i)$. Suppose there are two orthonormal basis $\{\ket{a_i}\}$ and $\{\ket{b_i}\}$ of $\cH_A$ such that $U=\sum_{i}(\proj{a_i}\ox I_B)U(\proj{b_i}\ox I_B)$. We assume the decomposition $U=\sum_{i,j}\ketbra{a_i}{b_j}\ox U_{ij}$, and obtain $U_{ij}=0$ for any $i\ne j$. So $U$ is a controlled unitary from the $A$ side, and $(ii)\ra(i)$ holds.

Next let us prove the relation $(iii)\ra(i)$. Suppose for two arbitrary orthonormal basis $\{\ket{c_i}\}$ and $\{\ket{e_j}\}$ of $\cH_B$, there are unitaries $S$ and $T$ on $\cH_A$ such that the operators $(S\ox\bra{c_i})U(T\ox\ket{e_j})$ on $\cH_A$, $\forall i,j$ are all diagonal. We can always have the decomposition $U=\sum_{i,j}U_{ij}\ox\ketbra{c_i}{e_j}$ with some operators $U_{ij}$ on $\cH_A$. Then all matrices $SU_{ij}T$ are diagonal. By definition $U$ is a controlled unitary from the $A$ side, and $(iii)\ra(i)$ holds. So we have shown that the first three assertions are equivalent.

To prove the claim in the last sentence of the lemma, suppose the operators $\{B_j\}$ are linearly independent. The relation $(iv)\ra(i)$ follows from the definition of controlled unitaries (see also \cite[Lemma 2]{cy13}). The relation $(v)\ra(i)$ is from \cite[Lemma 2]{cy14}.
We now show that $(i)\ra(iv),(v)$. Up to a permutation of labels, the $A$ space of $U$ is spanned by the linearly independent matrices $A_1,\cdots,A_r$. We have the Schmidt decomposition $U=\sum^r_{i=1} A_i \ox B_i'$, where each $B_i'$ is a linear combination of $\{B_j\}$. Since the operators $\{B_j\}$ are linearly independent, any $A_i$ is in the $A$ space of $U$, thus from the definition of controlled unitaries we have $(i)\ra(iv)$. It follows from \cite[Lemma 2]{cy14} that $(i)\ra(v)$ holds when $i,j=1,\cdots,r$ in assertion $(v)$. For any $k,l$, we have $A_kA_l^\dg\in\lin\{A_iA_j^\dg\}_{i,j=1,\cdots,r}$ and $A_k^\dg A_l\in\lin\{A_i^\dg A_j\}_{i,j=1,\cdots,r}$. It implies $(i)\ra(v)$, and thus the equivalence between any two of the five assertions $(i)-(v)$.
This completes the proof.
\epf

Note that the proof for $(iii)\ra(i)$ is similar to proving the so-called classical and generalized classical states \cite{ccm11}. The former has zero quantum discord, so there is no quantum correlation in the classical states \cite{oz01}. They are both a class of separable states lying near the quantum-classical boundary, in the sense that
each party of such states can perfectly identify a locally held state without disturbing the global state. This is a task known as non-disruptive local state identification, which is related to the problem of unambiguous state discrimination \cite{ccm11}. The counterpart of this task for controlled unitaries has been presented as Lemma~\ref{le:controlunitary} (ii). The generalized classical states also allow for another quantum-information task of \textit{local broadcasting} \cite{phh08}. An N-partite quantum state $\rho^{(12...N)}$ allows for local broadcasting if there exists local maps $\Lambda^{(i)}:\mathcal{H}^{(i)}\to\mathcal{H}^{(i)}\otimes\mathcal{H}^{(i)}$ such that the state $\sigma^{(11'22'...NN')}=[\Lambda^{(1)}\otimes...\otimes \Lambda^{(N)}]\rho$ has reduced states $\sigma^{(12...N)}=\sigma^{(1'2'...N')}=\rho$. However so far there is no counterpart of local broadcasting for controlled unitaries. Below we present a result that applies to all bipartite operators.

\bl
\label{le:nonsingular}
The following statements hold:

(i) The A space and the B space of a Schmidt-rank-$r$ bipartite operator each contains $r-1$ linearly independent singular operators.
\\
(ii) Suppose a Schmidt-rank-$r$ bipartite unitary with Schmidt decomposition $\sum^r_{i=1} A_i \ox B_i$ satisfies $B_i\ket{a}=0$ for a nonzero vector $\ket{a}$, $i=2,\cdots,r$. Then the unitary is a BCU from the $B$ side, and $A_1$ is proportional to a unitary matrix. That is, the $A$ space of $U$ contains a unitary matrix.
\el
\bpf
(i) Suppose a Schmidt expansion of the Schmidt-rank-$r$ bipartite unitary is $U=\sum_{j=1}^r A_j\ox B_j$. From Lemma~\ref{le:singular}, there is a nonzero linear combination of $A_1$ and $A_2$ that is singular, denoted as $A_1'$.  Apparently $A_1'$ is in the $A$ space of $U$. We can re-expand $U$ using $r$ linearly independent operators on the $A$ side including $A_1'$. Now the remaining $r-1$ A-side operators in such expansion also contains a singular operator, again from Lemma~\ref{le:singular}. Denote such operator as $A_2'$.  The $A_2'$ is linearly independent from $A_1'$, since the $r-1$ A-side operators above together with $A_1'$ form a linearly independent set. We can re-expand $U$ using $r$ operators on the $A$ side including $A_1'$ and $A_2'$. This process can repeat until we find $r-1$ linearly independent singular operators $A_1', A_2',\cdots,A_{r-1}'$. Hence the $A$ space of $U$ contains $r-1$ linearly independent singular operators. The proof is similar for the $B$ space.

(ii) Let $U=\sum^r_{i=1}A_i\ox B_i$. From $U^\dg U=I$, we have $\sum^r_{i,j=1} A_i^\dg A_j \ox B_i^\dg B_j = I$. Since $B_i\ket{a}=0$ for a nonzero vector $\ket{a}$, $i=2,\cdots,r$, we have $(I_A\ox\bra{a}) U^\dg U (I_A\ox\ket{a}) = \bra{a}B_1^\dg B_1 \ket{a} A_1^\dg A_1 = I_A$. So $A_1$ is proportional to a unitary matrix.

It remains to prove that $U$ is a BCU from the $B$ side.
Since $B_i\ket{a}=0$ for $i\ge 2$, there exist two $d_B\times d_B$ unitaries $V$ and $W$ such that $VB_1W=\left(\begin{array}{cc}
                     B_{11} & 0 \\
                     B_{12}  & B_{13} \\
                   \end{array}
                 \right)$ and
$VB_iW=\left(\begin{array}{cc}
                     B_{i1} & 0 \\
                     B_{i2} & 0 \\
                   \end{array}
                 \right),  \,\, i\ge 2$,
where $B_{11}$ and $B_{i1}$ are of size $(d_B-1)\times (d_B-1)$. Let $U'=(I_A\ox V)U(I_A\ox W)$, which is locally equivalent to $U$.
Since $U$ is unitary, $U'$ is also unitary and the complex number $B_{13}$ is nonzero. Since $(U')^\dg U'=I$, we have
\bea
\label{eq:u'}
\sum^r_{j=1} A_1^\dg A_j \ox B_{13}^* B_{j2}= 0.
\eea
We have proven that $A_1$ is proportional to a unitary and $B_{13}\ne0$. Thus \eqref{eq:u'} implies $\sum^r_{j=1} A_j \ox B_{j2}= 0$. Since $U=\sum^r_{j=1}A_j\ox B_j$ has Schmidt rank $r$, the $r$ matrices $A_1,\cdots,A_r$ are linearly independent. These arguments imply $B_{j2}=0$ for all $j$. So
\bea
U'&=&A_1\ox \left(\begin{array}{cc}
                     B_{11} & 0 \\
                     0  & B_{13} \\
                   \end{array}
                 \right)+\sum^r_{i=2}A_i\ox
\left(\begin{array}{cc}
                     B_{i1} & 0 \\
                     0 & 0 \\
                   \end{array}
                 \right)
\notag\\
&=&
\bigg(A_1\ox B_{13} \proj{d_B}\bigg)
+
\bigg(\sum^r_{i=1}A_i\ox  B_{i1} \bigg).
\eea
By definition $U'$ is a BCU from the $B$ side. Since $U$ and $U'$ are locally equivalent, $U$ is also a BCU from the $B$ side. This completes the proof.
\epf

%
%
%
%

\section{Schmidt-rank-three bipartite unitaries}
\label{sec:sch3}

In this section we show that any Schmidt-rank-three bipartite unitary is a controlled unitary. This is the main result, as proved in Theorem \ref{thm:sch3}. For this purpose, we present four preliminary lemmas \ref{le:BCUsch3}, \ref{le:nonB-BCU}, \ref{le:unitary_Aspace}, and \ref{le:1+(d-1)}. They characterize the properties of Schmidt-rank-three bipartite unitaries and some of them might be generalized to bipartite unitaries of higher Schmidt rank. The proof of these lemmas are intimately related to the linear algebra constructed in Sec. \ref{subsec:linearalg} and the concepts in the preliminaries such as BCU. Although they are both preliminaries for Theorem \ref{thm:sch3}, the lemmas in this section work for only Schmidt-rank-three bipartite unitaries.

\bl
\label{le:BCUsch3}
The following two statements are equivalent:\\
(i) any Schmidt-rank-three bipartite unitary is a controlled unitary;\\
(ii) any Schmidt-rank-three bipartite unitary is a BCU.
\el
\bpf
To prove the lemma, we first prove the following statement:\\
\textit{If a bipartite unitary $U$ of Schmidt rank three is the direct sum from A side of two unitaries $T_1,T_2$, and satisfies one of the following conditions:\\
(a) one of $T_1$, $T_2$ is a Schmidt-rank-three controlled unitary from B side;\\
(b) both $T_1$, $T_2$ are controlled unitaries;\\
then $U$ is a controlled unitary.}

The proof for the statement is as follows:

Suppose $U=T_1 \oplus_A T_2$, where $T_1,T_2$ are unitaries.

For condition (a): without loss of generality assume $T_1$ is controlled from the $B$ side and has Schmidt rank three. From Lemma~\ref{le:controlunitary}, the basis of three $B$ side operators for $T_1$ can be simultaneously diagonalized under local unitaries. Since $U$ is also of Schmidt rank three, the linear span of such three $B$ side operators contains the $B$ side operators in the Schmidt expansion of $T_2$. Hence all $B$ side operators of $U$ are simultaneously diagonalized under local unitaries. From Lemma~\ref{le:controlunitary}, $U$ is a controlled unitary from the $B$ side. So the assertion follows.

For condition (b): If $T_1,T_2$ are both controlled from the $A$ side, then $U$ is also a controlled unitary from the $A$ side. Otherwise, it must be that one of the two controlled unitaries, say $T_1$ is controlled from the $B$ side only. Then $T_1$ is of Schmidt rank at least 3, since unitaries with Schmidt rank not exceeding 2 are controlled from both sides \cite{cy13}. Since $U$ has Schmidt rank three, so is $T_1$. Thus condition (i) holds, and hence $U$ is a controlled unitary from the $B$ side.  This completes the proof of the statement.

For the lemma, the implication $(i)\ra(ii)$ is obvious. To prove the implication $(ii)\ra(i)$, we present a method for proving that any Schmidt-rank-three bipartite unitary is a controlled unitary that involves induction over the dimensions $d_A$ and $d_B$. The induction hypothesis is that any Schmidt-rank-three bipartite unitary on $k\times l$ ($k\ge 2$, $l\ge 2$) system is a controlled unitary, when either $k\le d_A-1$, $l\le d_B$ or $k\le d_A$, $l\le d_B-1$. The boundary case is that $k=2$ or $l=2$, and is proved in \cite{cy14}. By condition (ii), the Schmidt-rank-three unitaries $U$ on $d_A\times d_B$ system is always a BCU. We can always choose to divide $U$ into two blocks only rather than many blocks, and use the induction hypothesis to get that each block is a controlled unitary, then the previously proved statement when condition (b) holds implies that the whole unitary is a controlled unitary. This proves the incremental case in the induction hypothesis when the dimensions are ($d_A$, $d_B$). This completes the proof.
\epf

\bl
\label{le:nonB-BCU}
Consider a Schmidt-rank-three bipartite unitary operator whose A space is spanned by unitary matrices. Then the operator is $\sum^3_{i=1}A_i\ox B_i$, where $\{A_i\}$ are unitary and at least one of the $B_i$ is non-invertible.
\el
\bpf
Suppose the Schmidt-rank-three bipartite unitary is $U=\sum^3_{i=1}X_i\ox Y_i$ where $\{X_i\}$ are unitary. If one of $\{Y_i\}$ is non-invertible then the assertion follows. It suffices to consider that $\{Y_i\}$ are all invertible. There is a complex number $x$ such that $B_1=Y_1-xY_2$ is non-invertible. We have
$U = X_1 \ox B_1 + (xX_1+X_2)\ox Y_2 + X_3 \ox Y_3$. Let $\ket{a}$ be a state such that $B_1\ket{a}=0$.
Since $(I_A \ox \bra{a})U^\dg U(I_A \ox \ket{a}) =I_A$, we have
\bea
\label{eq:nonB-BCU}
\bigg( (x X_1+X_2)\ox Y_2\ket{a} + X_3 \ox Y_3\ket{a} \bigg)^\dg \cdot \bigg( (x X_1+X_2)\ox Y_2\ket{a} + X_3 \ox Y_3\ket{a} \bigg) = I_A.
\eea
Because $Y_2$ and $Y_3$ are invertible, we have $Y_2\ket{a}\ne0$ and $Y_3\ket{a}\ne0$. Thus there is a complex number $y$ such that
$-yY_2\ket{a}+Y_3\ket{a}$ is orthogonal to $Y_2\ket{a}$. Since $X_3$ is unitary, Eq. \eqref{eq:nonB-BCU} implies that $xX_1+X_2+yX_3$ is nonzero and proportional to a unitary. We have $
U = X_1 \ox B_1 + (x X_1+X_2+y X_3)\ox Y_2 + X_3 \ox (-y Y_2+Y_3).
$
By assuming $X_1=A_1$, the second and third products as $A_2\ox B_2$ and $A_3\ox B_3$, respectively, we obtain the assertion.
This completes the proof.
\epf

\bl
\label{le:unitary_Aspace}
Any Schmidt-rank-three non-BCU bipartite unitary cannot have one of its spaces spanned by unitary matrices.
\el
\bpf
The proof is by contradiction. Assume there is a Schmidt-rank-three non-BCU bipartite unitary satisfying that one of its spaces is spanned by unitary matrices.  Let $U$ be such a Schmidt-rank-three non-BCU bipartite unitary whose $A$ space is spanned by three unitary matrices. These three unitaries are not simultaneously diagonalizable, since otherwise $U$ is a controlled unitary, contradicting with the non-BCU condition. From Lemma \ref{le:nonB-BCU}, we have $U=\sum_{j=1}^3 A_j\ox B_j$, where $A_j$ are all unitary and $B_3$ is non-invertible. Let $B_3$ be of rank $r\in [1,d_B-1]$. Up to local unitaries on $\cH_B$, we may assume that $B_3=P\op 0_{d_B-r}$ is diagonal and $P$ is a $r\times r$ invertible matrix. Let
$B_i=\left(
                   \begin{array}{cc}
                     B_{i1} & B_{i2} \\
                     B_{i3} & B_{i4}
                   \end{array}
                 \right) $ for $i=1,2$, and $B_{i1}$ is of size $r\times r$. If $B_{12}=B_{22}=0$, then there are two complex numbers $x,y$ such that $xB_{14}+yB_{24}$ is nonzero and singular. We re-expand the unitary as $U=A_1' \ox (xB_1+yB_2) + A_2' \ox B_2' + A_3 \ox B_3$. Let $\ket{a}\in\lin\{\ket{r+1},\cdots,\ket{d_B}\}$ be a state such that $(xB_{14}+yB_{24})\ket{a}=0$. Hence $(xB_1+yB_2)\ket{a}=B_3\ket{a}=0$. Lemma \ref{le:nonsingular} (ii) implies that $U$ is a BCU from the $B$ side, a contradiction with the assumption. Below we assume that one of $B_{12}$ and $B_{22}$ is not zero.

Up to a unitary similarity transform on $\cH_A$, we may assume that $A_1=I_A$, $A_2=D$, and the identical diagonal entries of $D$ are adjacent. Let $D=e^{i\t_1} P_1 + e^{i\t_2} P_2 + \cdots + e^{i\t_k} P_k$, where $k\ge2$, the projectors $P_i$ are pairwise orthogonal, and $\t_i\in(0,2\p)$ are pairwise different.
We write $A_3$ as a partitioned matrix $A_3=[V_{ij}]_{i,j=1,\cdots,k}$, where the submatrix $V_{ij}$ is of size $\abs{P_i}\times \abs{P_j}$, and the diagonal blocks $V_{ii}$ are all upper-triangular. If the blocks $V_{jm}$ are zero for all distinct integers $j, m$, then the unitarity of $U$ implies that all $V_{ii}$ are diagonal. $U$ is a controlled unitary from the $A$ side, it is a contradiction with the assumption that the A-side unitaries are not simultaneously diagonalizable. So there are two distinct integers $j,m$ such that $V_{jm}\ne0$. Since $U$ is unitary, there is another integer $n$ different from $j$, such that $V_{nj}\ne0$.
Let $V_{ii}'$ be the same as $V_{ii}$, except that the diagonals of $V_{ii}'$ are replaced by zeros. From $U^\dg U=I$ we have
\bea
\label{eq:onespaceunitary}
&&
I_A \ox (B_1^\dg B_1 + B_2^\dg B_2 + B_3^\dg B_3)
+
D \ox B_1^\dg B_2
\notag\\
&+&
D^\dg \ox B_2^\dg B_1
+
A_3 \ox B_1^\dg B_3
+
A_3^\dg \ox B_3^\dg B_1
\notag\\
&+&
D^\dg A_3 \ox B_2^\dg B_3
+
A_3^\dg D \ox B_3^\dg B_2
= I.
\eea
By considering the off-diagonal part of the $A$ side operators in this equation, we obtain that the following partitioned matrix
\begin{align}
\label{eq:partition}
&
\left(
                   \begin{array}{ccc}
                     V_{nn}' & V_{nj} & V_{nm} \\
                     V_{jn} & V_{jj}' & V_{jm} \\
                     V_{mn} & V_{mj} & V_{mm}' \\
                   \end{array}
                 \right)
                 \ox B_1^\dg B_3
+
\notag\\
&
\left(
                   \begin{array}{ccc}
                     (V_{nn}')^\dg & V_{jn}^\dg & V_{mn}^\dg \\
                     V_{nj}^\dg & (V_{jj}')^\dg & V_{mj}^\dg \\
                     V_{nm}^\dg & V_{jm}^\dg & (V_{mm}')^\dg \\
                   \end{array}
                 \right)
                 \ox B_3^\dg B_1
+
\notag\\
&
\left(
                   \begin{array}{ccc}
                     e^{-i \t_n} V_{nn}' & e^{-i \t_n}V_{nj} & e^{-i \t_n}V_{nm} \\
                     e^{-i \t_j} V_{jn} & e^{-i \t_j}V_{jj}' & e^{-i \t_j}V_{jm} \\
                     e^{-i \t_m}V_{mn} & e^{-i \t_m}V_{mj} & e^{-i \t_m}V_{mm}' \\
                   \end{array}
                 \right) \ox B_2^\dg B_3
+
\notag\\
&
\left(
                   \begin{array}{ccc}
                     e^{i \t_n} (V_{nn}')^\dg & e^{i \t_j}V_{jn}^\dg & e^{i \t_m}V_{mn}^\dg \\
                     e^{i \t_n}V_{nj}^\dg & e^{i \t_j}(V_{jj}')^\dg & e^{i \t_m}V_{mj}^\dg \\
                     e^{i \t_n}V_{nm}^\dg & e^{i \t_j}V_{jm}^\dg & e^{i \t_m}(V_{mm}')^\dg \\
                   \end{array}
                 \right)   \ox B_3^\dg B_2
\end{align}
is zero. In particular if $m=n$, then each of the four blocks in the $A$ side of \eqref{eq:partition} is replaced by its upper left four blocks, say $\left(
                   \begin{array}{ccc}
                     V_{nn}' & V_{nj} & V_{nm} \\
                     V_{jn} & V_{jj}' & V_{jm} \\
                     V_{mn} & V_{mj} & V_{mm}' \\
                   \end{array}
                 \right)$ is replaced by
$\left(
                   \begin{array}{cc}
                     V_{nn}' & V_{nj} \\
                     V_{jn} & V_{jj}'
                                       \end{array}
                 \right)$, etc.
Since $V_{jm}$ and $V_{nj}$ are nonzero and $\t_j\in (0,2\p)$ are pairwise different, the space spanned by the four operators in the $A$ side of \eqref{eq:partition} has dimension at least two. From \eqref{eq:partition} and Lemma~\ref{le:schineq}, the space $H$ spanned by the four operators $B_1^\dg B_3$, $B_3^\dg B_1$, $B_2^\dg B_3$, and $B_3^\dg B_2$ has dimension at most two. In the second paragraph we have shown that $P$ is a $r\times r$ invertible matrix, and one of $B_{12}$ and $B_{22}$ is not zero. It implies $\dim H=2$. By using this fact and also the block structure of the four B-side matrices, it can be determined that $B_1^\dg B_3\propto B_2^\dg B_3$, and $B_3^\dg B_1\propto B_3^\dg B_2$.  Because $\dim H=2$ and none of the four operators in $A$ side of \eqref{eq:partition} is zero, neither are the operators $B_1^\dg B_3$, $B_3^\dg B_1$, $B_2^\dg B_3$, and $B_3^\dg B_2$.
The two matrices $B_1^\dg B_3$ and $B_3^\dg B_1$ are linearly independent. Applying to \eqref{eq:partition} these facts and that $V_{jm}$ and $V_{nj}$ are nonzero, we have $\t_j=\t_n$. It is a contradiction with the fact that $\t_j,\t_n\in (0,2\p)$ are different. Therefore the assumption is wrong.
This completes the proof.
\epf

\bl
\label{le:1+(d-1)}
Any Schmidt-rank-three non-BCU bipartite unitary $U=\sum^3_{j=1}A_j\ox B_j$ cannot satisfy the following condition:
\\
$A_1$ is the direct sum of a $2\times2$ upper left diagonal matrix and a $(d_A-2)\times (d_A-2)$ matrix, and $A_2$ is the direct sum of another upper left $2\times2$ diagonal matrix and another $(d_A-2)\times (d_A-2)$ matrix.  The diagonal vectors of the two $2\times2$ matrices are linearly independent.
\el
\bpf
The proof is by contradiction. Assume the condition stated in the lemma holds. By replacing $A_1,A_2$ by a suitable linear combination of them and absorbing two global factors into $B_1,B_2$, we may assume that the first two diagonal entries of $A_1$ and $A_2$ are $1,0$ and $0,1$, respectively. By replacing $A_3$ by a suitable linear combination of $A_1,A_2$, and $A_3$, we may assume that the first two diagonal entries of $A_3$ are zero. So the first rows of $A_1$ and $A_3$ are orthogonal, and the first columns of $A_1$ and $A_3$ are also orthogonal. Up to local unitaries on $\cH_B$, we may assume that $B_3=\sum^k_{i=1} c_i P_i$, where the projectors $P_i$ are pairwise orthogonal, and the real and nonnegative $c_i$ are pairwise different. Since $(\bra{1}\ox I_B)U^\dg U(\ket{1}\ox I_B)=(\bra{1}\ox I_B)U U^\dg(\ket{1}\ox I_B)=I_B$, we have
\bea
\label{eq:twodiagonal1}
B_1^\dg B_1 + w B_3^\dg B_3
=
B_1 B_1^\dg + x B_3 B_3^\dg
=
I_B,
\eea
with nonnegative and real numbers $w,x$. Since $(\bra{2}\ox I_B)U^\dg U(\ket{2}\ox I_B)=(\bra{2}\ox I_B)U U^\dg(\ket{2}\ox I_B)=I_B$, we have
\bea
\label{eq:twodiagonal2}
B_2^\dg B_2 + y B_3^\dg B_3
=
B_2 B_2^\dg + z B_3 B_3^\dg
=
I_B,
\eea
with nonnegative and real numbers $y,z$. If one of $w,x,y,z$ is zero, then Lemma~\ref{le:nonsingular} implies that $U$ is a BCU, a contradiction with the non-BCU condition in the lemma.  Below we assume that none of $w,x,y,z$ is zero. From $B_3=\sum^k_{i=1} c_i P_i$, Eqs. \eqref{eq:twodiagonal1} and \eqref{eq:twodiagonal2}, we have $w=x$ and $y=z$. Thus $B_1,B_2$ are normal. We have $B_1B_1^\dg=\sum^k_{i=1} (1-wc_i^2) P_i$, and $B_2B_2^\dg=\sum^k_{i=1} (1-yc_i^2) P_i$. Since $\{c_i\}$ are real, nonnegative, and pairwise different, so are $\{1-wc_i^2\}$ and $\{1-yc_i^2\}$, respectively. Then Lemma \ref{le:normal} implies that $B_1=\op^k_{i=1} (1-wc_i^2)^{\frac12} U_i$ and $B_2=\op^k_{i=1} (1-yc_i^2)^{\frac12} V_i$, where $U_i,V_i$ are unitary matrices on the subspace of projector $P_i$ and $U_i^\dg U_i = V_i^\dg V_i = P_i$ for all $i$. So $U$ is the B-direct sum of the bipartite unitaries $X_i=A_1 \ox (1-wc_i^2)^{\frac12} U_i + A_2 \ox (1-yc_i^2)^{\frac12} V_i + A_3 \ox c_i P_i$, $i=1,\cdots,k$.
If $k>1$, $U$ is a BCU from the $B$ side, a contradiction with the assumption in the lemma. Hence $k=1$.  Since $U$ has Schmidt rank three, the $B$ space of $U$ is spanned by three unitary matrices $U_1, V_1, I_B$. This is a contradiction with Lemma~\ref{le:unitary_Aspace}, given that $U$ is non-BCU. Hence the assumption is wrong. This completes the proof.
\epf

\bt
\label{thm:sch3}
Any bipartite unitary of Schmidt rank three is a controlled unitary.
\et
\bpf
Let $U=\sum^3_{j=1}A_j\ox B_j$ be a Schmidt-rank-three bipartite unitary. By Lemma~\ref{le:BCUsch3}, we only need to consider the cases that $U$ is not a BCU, for the following reason: if we always get a contradiction for the non-BCU cases, for all dimensions $d_A,d_B$, then we would have proved that any Schmidt-rank-three bipartite unitary is a BCU, and from Lemma~\ref{le:BCUsch3}, any bipartite unitary of Schmidt rank three is a controlled unitary. Hence in the following we assume $U$ is not a BCU. From Lemma \ref{le:nonsingular}(i) we may assume that $A_1$ is singular. Up to local unitaries on $\cH_A$, we may assume $A_1\ket{1}=A_1^\dg\ket{1}=0$. Since $(\bra{1}\ox I_B)U^\dg U(\ket{1}\ox I_B)=(\bra{1}\ox I_B)UU^\dg(\ket{1}\ox I_B)=I_B$, we have
\bea\label{eq:sch31}
I_B=x_1 B_2^\dg B_2 + y_1 B_3^\dg B_3 + z_1 B_2^\dg B_3 + z_1^\ast B_3^\dg B_2,
\notag\\
I_B=x_2 B_2 B_2^\dg + y_2 B_3 B_3^\dg + z_2 B_2 B_3^\dg + z_2^\ast B_3 B_2^\dg,
\eea
where $x_i,y_i$ are non-negative and real numbers, $z_i$ are complex numbers. They are given by $x_1=\bra{1}A_2^\dg A_2\ket{1}$, $y_1=\bra{1}A_3^\dg A_3\ket{1}$, $z_1=\bra{1}A_2^\dg A_3\ket{1}$, $x_2=\bra{1}A_2 A_2^\dg\ket{1}$, $y_2=\bra{1}A_3 A_3^\dg\ket{1}$, $z_2=\bra{1}A_2 A_3^\dg\ket{1}$. From these equations we get that the inequalities $x_iy_i\ge \abs{z_i}^2$ hold for $i=1,2$. If one of the equalities hold then either $T\ket{1}=0$ or $T^\dg\ket{1}=0$ for some $T\in\lin\{A_2,A_3\}$, then from Lemma~\ref{le:nonsingular}(ii), $U$ is a BCU, and we get a contradiction. So $x_iy_i> \abs{z_i}^2$ for $i=1,2$. It follows from Eq.~\eqref{eq:sch31} and Lemma \ref{le:orthogonalization} that there are two linearly independent matrices $B_4,B_5$ in $\lin\{B_2,B_3\}$ such that $B_4,B_5$ are simultaneously diagonalizable. Hence $U$ is locally equivalent to the non-BCU bipartite unitary satisfying the condition stated in Lemma~\ref{le:1+(d-1)}, but that lemma says the condition cannot be satisfied for non-BCU Schmidt-rank-three unitaries. This is a contradiction, hence we have shown a contradiction always exists for any dimensions $d_A,d_B$. Hence there is no non-BCU Schmidt-rank-three unitary in systems of any size, meaning that any Schmidt-rank-three bipartite unitary is a BCU. Then by Lemma~\ref{le:BCUsch3}, any bipartite unitary of Schmidt rank three is a controlled unitary. This completes the proof.
\epf

We can further
decide the side from which $U$ is controlled by Lemma
\ref{le:controlunitary}, and the algorithm is described in \cite{mm11}.

\section{Applications}
\label{sec:app}

In this section we propose a few applications of our results on
general nonlocal unitary operators. To generalize our findings to multipartite case, we study Schmidt-rank-three multipartite unitary operators in Sec. \ref{sec:multi}. We begin by defining the necessary terminologies, and then characterize the basic properties of multipartite unitaries in Lemma \ref{le:multiunitary}. We also present a few methods of constructing non-controlled Schmidt-rank three multipartite unitaries in Example \ref{ex:noncontrol}. The main result of this section is that any multipartite unitary operator of Schmidt rank three can be controlled by one system or collectively controlled by two systems, regardless of the number of systems of this operator in Theorem \ref{thm:multi}. This is based on the preliminary lemmas \ref{le:noninvertible}, \ref{le:nonzerolinearcombi} and \ref{le:sr3=kcollective}. The theorem is further strengthened in Corollary \ref{cr:multi}, by which we show that every Schmidt-rank-three multipartite unitary is controlled by the union of two systems.

Next, we study the Schmidt-rank-three $n$-qubit unitary $U$ in Sec. \ref{subsect:multiqubit}. We construct non-controlled $U$ for any odd $n\ge3$ in \eqref{eq:sr3nqubit}. The main result of this section is Proposition \ref{pp:evenqubit}, which states that $U$ for any even $n\ge4$ is a controlled unitary. This is based on a preliminary lemma \ref{le:44}. In Sec. \ref{sec:entanglementcost}, we show that any bipartite unitary operator of Schmidt rank three can be implemented by
LOCC and a maximally entangled state $\ket{\Ps_r}={1\over \sqrt
r}\sum^r_{i=1}\ket{ii}$, where $r=\min\{d_A^2,d_B\}$ and $d_A\le d_B$, see Lemma \ref{le:costSR3}. In Sec. \ref{sec:schmidtnumber}, we study the connection between the controlled unitary and Schmidt rank in Lemma \ref{le:schnum}. In Sec. \ref{sec:cy13}, we give an alternative proof of one of the main results in \cite{cy13}, using only the linear algebra developed in Sec. \ref{sec:pre}. The main result is Lemma \ref{le:sch2}.

\subsection{Schmidt-rank-three multipartite unitary operators}
\label{sec:multi}

Multipartite quantum states, such as the Greenberger-Horne-Zeilinger (GHZ) states, W states \cite{dvc2000} and graph states
\cite{br01}, are the fundamental ingredients of quantum information processing. Multipartite quantum states can be generated by using multipartite unitary operations and in particular, by using multipartite controlled unitary operations. Understanding the structure of multipartite unitary helps explore the problems such as the state classification,  implementation, and the experimental violation of
multipartite Bell-type inequalities \cite{lzj14}. In this section, we will investigate multipartite unitary operators of Schmidt rank three, by using the results developed for bipartite unitaries.

To investigate multipartite unitary operators, we generalize the definitions and terminologies for bipartite unitary operators. Let $j_1,\cdots,j_k$ be $k$ distinct integers in the set $\{1,2,\cdots,n\}$, and $\overline{j_1,\cdots,j_k}$ the remaining $n-k$ distinct integers in the same set. We shall denote the $k$ systems $\cA_{j_1},\cdots,\cA_{j_k}$ and the corresponding Hilbert space $\cH_{j_1}\ox\cdots\ox \cH_{j_k}$ as $\cA_{j_1,\cdots,j_k}$ and $\cH_{j_1,\cdots,j_k}$, respectively. Let $U$ be an $n$-partite unitary operator of the system $\cA_{1,\cdots,n}$ in the space $\cH_{1,\cdots,n}$ with $\dim \cH_i=d_i$ for all $i$. We say that $U$ has Schmidt rank $r$ if $U=\sum^r_{i=1} A_{i,1}\ox A_{i,2} \ox \cdots \ox A_{i,n}$, and $U$ cannot be the sum of fewer product operators. $U$ is a controlled unitary from the $\cA_j$ side when it is locally equivalent to $\sum^{d_j}_{j=1}\proj{j} \ox V_j$ where $V_j$ is a unitary on the subspace $\cH_{\overline{j}}$. We shall sometimes ignore the subscript $j$ and just say that $U$ is a controlled unitary.  We say $U$ is not a controlled unitary, or equivalently, $U$ is a ``non-controlled'' unitary, when $U$ is not controlled from any one system, but could be collectively controlled from two or more systems.

If we partition the system $\cA_{1,\cdots,n}$ into two larger systems $\cA_{j_1,\cdots,j_k}$ and $\cA_{\overline{j_1,\cdots,j_k}}$, then we can regard $U$ as a bipartite unitary $U_{j_1,\cdots,j_k:\overline{j_1,\cdots,j_k}}$. Its Schmidt rank is defined as the dimension of the $\cA_{j_1,\cdots,j_k}$ space of $U$. It is evidently equal to the dimension of the $\cA_{\overline{j_1,\cdots,j_k}}$ space. Since $U$ has Schmidt rank $r$, the Schmidt rank of $U_{j_1,\cdots,j_k:\overline{j_1,\cdots,j_k}}$ is between $1$ and $r$. In particular, if it is equal to one, then $U$ is the tensor product of two multipartite unitary operators on the subspaces $\cH_{j_1,\cdots,j_k}$ and $\cH_{\overline{j_1,\cdots,j_k}}$, respectively. Based on these terminologies and definitions, we characterize the properties of multipartite unitary as follows.

\bl
\label{le:multiunitary}
Let $U=\sum^r_{i=1} A_{i,1}\ox A_{i,2} \ox \cdots \ox A_{i,n}$ be an $n$-partite Schmidt-rank-$r$ unitary operator. Then
\\
(i) For any integer $j\in [1,n]$,  the $r$ product operators $\{A_{i,1} \ox \cdots \ox A_{i,j-1} \ox A_{i,j+1} \ox \cdots \ox A_{i,n}\}_{i=1,\cdots,r}$ are linearly independent.
\\
(ii) $U$ is a controlled unitary from the $\cA_j$ side if and only if
the $r$ operators $A_{1,j},\cdots,A_{r,j}$ have simultaneous singular value decomposition.
\\
(iii) $U$ is a controlled unitary if there are integers $j_1,\cdots,j_k$ such that the $r$ product operators $\{A_{i,j_1}\ox\cdots\ox A_{i,j_k}\}_{i=1,\cdots,r}$ span a 1-dimensional or 2-dimensional space.
\el
\bpf
(i) Assume the assertion does not hold. Then one of these $r$ product operators is a linear combination of the other $r-1$ product operators. By expanding $U$ using such $r-1$ operators that act on the space $\cH_{1,\cdots,j-1,j+1,\cdots,n}$, we obtain an expansion of $U$ with $r-1$ terms which are product operators.  Thus $U$ has Schmidt rank not greater than $r-1$, a contradiction with the assumption. Hence the assertion holds.

(ii) From (i) and the fact that $U$ has Schmidt rank $r$, the $r$ product operators $\{A_{i,1} \ox \cdots \ox A_{i,j-1} \ox A_{i,j+1} \ox \cdots \ox A_{i,n}\}_{i=1,\cdots,r}$ are linearly independent. The assertion follows from Lemma \ref{le:controlunitary} (iv).

(iii) Under the first condition, $U$ is a controlled unitary from any system of $\cA_{j_1,\cdots,j_k}$. So the assertion holds. It suffices to prove the assertion under the second condition. Suppose the $r$ product operators $\{A_{i,j_1}\ox\cdots\ox A_{i,j_k}\}_{i=1,\cdots,r}$ span a 2-dimensional space. Without loss of generality, let $A_{1,j_1}\ox \cdots \ox A_{1,j_k}$ and $A_{2,j_1}\ox \cdots \ox A_{2,j_k}$ be linearly independent. Thus $A_{l,j_1}\ox \cdots \ox A_{l,j_k}$ with any $l>2$ is their linear combination. By the same reason, there is at least one integer, say $j_1$ such that $A_{1,j_1}$ and $A_{2,j_1}$ are linearly independent. So $A_{l,j_1}$ with any $l>2$ is their linear combination. By partitioning the system $\cA_{1,\cdots,n}$ into $\cA_{j_1}$ and $\cA_{\overline{j_1}}$, we obtain a Schmidt-rank-2 bipartite unitary $U_{j_1:\overline{j_1}}$. It is a controlled unitary from the $\cA_{j_1}$ side \cite{cy13}, and so is $U$. This completes the proof.
\epf

The ``if'' condition in Lemma \ref{le:multiunitary} (iii) means that the $\cA_{j_1,\cdots,j_k}$ space has dimension one or two, but the converse is generally not true. So the corollary cannot be generalized to multipartite unitary of a larger Schmidt rank, as we show in the following example.

\bex
\label{ex:4qubitunitary}
{\rm
Let $V$ be the two-qubit SWAP gate, $W=(\s_0\ox\s_3)V(\s_0\ox\s_3)$, and $U={1\over\sqrt2} (V_{12} \ox V_{34} + i W_{12} \ox W_{34})$. One can easily verify that $U$ is a four-qubit unitary operator of Schmidt rank bigger than three. The $\cA_{jk}$ space of $U$ has dimension equal to or larger than two, and the equality holds when $(j,k)=(1,2)$. Furthermore, the $\cA_j$ space with any $j$ has dimension four.  So $U$ is not a controlled unitary.
\qed
}
\eex

One may develop more conditions equivalent to Lemma \ref{le:multiunitary} (ii), by using the items in Lemma \ref{le:controlunitary}. It is known that any Schmidt-rank-2 multipartite unitary is a controlled unitary from any side \cite{cy13}. For $n=2$, we have shown that all Schmidt-rank-three unitaries are controlled unitaries in \cite{cy14} and Theorem \ref{thm:sch3}. Below we present a few examples and methods of constructing non-controlled Schmidt-rank-three multipartite unitary.

\bex
\label{ex:noncontrol}
{\rm
For $n=3$, there has been a Schmidt-rank-three non-controlled three-qubit unitary \cite{cy13}
\bea
\label{eq:sr3threequbit}
U^{(3)} = {1\over\sqrt3} (\s_0\ox \s_0 \ox \s_0 + i\s_1 \ox \s_1 \ox \s_1+ i\s_3 \ox \s_3 \ox \s_3 ).
\eea
Below we construct two other sets of Schmidt-rank-three non-controlled unitaries of high dimensions. It is easy to verify that the matrix $U+\s_0\ox \s_0 \ox (\sum^n_{i=3}\proj{i})$ is a $2\times2\times n$ Schmidt-rank-three non-controlled unitary. Next, Let $V=U\ox I_{\cA_1' \cA_2' \cA_3'}$ be a tripartite unitary of the systems $\cB_1,\cB_2,\cB_3$, where $\cB_i=\cA_i\cA_i'$ and $I_{\cA_i'}$ acts on the space $\bC^{d_i'}$ for $i=1,2,3$. Then $V$ is a Schmidt-rank-three non-controlled unitary on the $2d_1' \times 2d_2' \times 2d_3'$ system. \qed
}
\eex

In spite of these examples, we do not have a systematic method of constructing Schmidt-rank-three non-controlled unitaries.

%
%

Although the tripartite unitary $U^{(3)}$ is not a controlled unitary, the bipartite unitary $U^{(3)}_{j_1,j_2:\overline{j_1,j_2}}$ with any distinct $j_1,j_2$ is a controlled unitary by  Theorem \ref{thm:sch3} and Lemma \ref{le:multiunitary} (ii). That is, $U^{(3)}$ is collectively controlled by system $\cA_{12}$. Next, the four-qubit unitary in Example \ref{ex:4qubitunitary} is a unitary collectively controlled by the system $\cA_{12}$ because of \cite{cy13}. These facts turn out to be general. In Theorem \ref{thm:multi} we will show that any Schmidt-rank-three multipartite unitary is collectively controlled by at most two systems. To prove this theorem, we present three preliminary lemmas.

\bl
\label{le:noninvertible}
Let $A_1,A_2,B_1,B_2$ be square matrices, and $A_1$ singular. The following statements hold:
\\
(i) If $B_2$ is singular, then the span of $A_1 \ox B_1$ and $A_2\ox B_2$ does not contain any invertible matrix.
\\
(ii) If $A_1 \ox B_1 + A_2\ox B_2$ is unitary, then $B_2$ is proportional to a unitary matrix.
\el
\bpf
(i) Since $A_1,B_2$ are both singular, let $\ket{a},\ket{b}$ be the states such that $A_1\ket{a}=B_2\ket{b}=0$. Then any matrix $V$ in the span of $A_1 \ox B_1$ and $A_2\ox B_2$ satisfies $V\ket{a,b}=0$. So $V$ is singular and the assertion holds.

(ii) Let $U=A_1 \ox B_1 + A_2\ox B_2$ be the unitary. Since $A_1$ is singular, let $\ket{a}$ be the state such that $A_1\ket{a}=0$. Then $(\bra{a}\ox I_B)U^\dg U(\ket{a}\ox I_B)=I_B\propto B_2^\dg B_2$. Hence $B_2$ is proportional to a unitary matrix. This completes the proof.
\epf

\bl
\label{le:nonzerolinearcombi}
Let $\sum^3_{i=1} A_{i,1}\ox A_{i,2} \ox A_{i,3}$ be a Schmidt-rank-three tripartite unitary. The matrices in one of the two sets $\{A_{i,2}\}_{i=1,2,3}$ and $\{A_{i,3}\}_{i=1,2,3}$ have simultaneous singular value decomposition, if one of the following two conditions is satisfied:
\\
(i) there is a state $\ket{b}$ such that any two vectors in $\{A_{i,1}\ket{b}\}_{i=1,2,3}$ are nonzero and parallel.
\\
(ii) The matrices in $\{A_{i,1}\}_{i=1,2,3}$ have simultaneous singular value decomposition, and two matrices in $\{A_{i,1}\}_{i=1,2,3}$ are invertible.
\el
\bpf
(i) Let $U=\sum^3_{i=1} A_{i,1}\ox A_{i,2} \ox A_{i,3}$. Since $U$ is unitary, we have $(\bra{b}\ox I_{\cA_{2,3}}) U^\dg U (\ket{b}\ox I_{\cA_{2,3}})=I_{\cA_{2,3}}=V^\dg V$, where $V=\sum^3_{i=1} x_i A_{i,2} \ox A_{i,3}$ with nonzero complex numbers $x_1,x_2,x_3$. So $V$ is unitary. Since $U$ has Schmidt rank three, so is $V$ by Lemma \ref{le:multiunitary} (i).
Theorem \ref{thm:sch3} implies that $V$ is a controlled unitary.
The assertion follows from Lemma \ref{le:controlunitary} (iv).

(ii) If $U$ satisfies condition (ii), then it satisfies condition (i). So the assertion follows. This completes the proof.
\epf

\bl
\label{le:sr3=kcollective}
Let $U=\sum^3_{i=1} A_{i,1}\ox A_{i,2} \ox \cdots \ox A_{i,n}$ be an $n$-partite Schmidt-rank-$3$ unitary operator collectively controlled by $k\le \lc {n \over 2} \rc$ systems of $\cA_{1,\cdots,n}$. Then
\\
(i) for any integers $j_1,\cdots,j_s\in [1,n]$ and $s\in[1,k-1]$, the $\cA_{j_1,\cdots,j_s}$ space of $U$ is spanned by unitary matrices.
\\
\\
(ii) If $k\ge 2$, then
\\
(ii.a) for any integer $j \in [1,n]$, the space $\lin\{A_{i,j}\}_{i=1}^3$ is spanned by three unitary matrices.
\\
(ii.b)
$U$ is a Schmidt-rank-three bipartite unitary on any bipartite cut.
\\
\\
(iii) If $k\ge 3$, then
\\
(iii.a)
there are $k$ distinct integers $l=j_1,\cdots,j_k \in [1,n]$ such that for each $l$, the set $\{A_{i,l}\}_{i=1,2,3}$ contains a singular matrix.
\\
(iii.b) Let $\{A_{i_l,l}\}_{i_l\in[1,3],l\in[1,n]}$ be the set of all singular matrices in $\{A_{i,j}\}$. Then the set $\{i_l\}$ consists of exactly two distinct integers in $\{1,2,3\}$. If $x\in\{1,2,3\}\sm\{i_l\}$, then $A_{x,l}$ with any $l\in[1,n]$ is proportional to a unitary matrix.
\\
(iii.c)
Any matrix $A_{i,j}$ is normal.
\el
\bpf
(i) If the $\cA_{j_1,\cdots,j_s}$ space of $U$ has dimension one or two, then it is spanned by unitary matrices \cite{cy13}. Suppose the space has dimension three. Theorem \ref{thm:sch3} implies that $U_{j_1,\cdots,j_s:\overline{j_1,\cdots,j_s}}$ is a controlled unitary. Since $s<k$, the system $\cA_{j_1,\cdots,j_s}$ cannot control $U$. So the $\cA_{j_1,\cdots,j_s}$ space of $U$ is spanned by unitary matrices. The assertion follows.

(ii) Let us prove (ii.a). Since $k\ge 2$, the unitary $U_{j_1:\overline{j_1}}$ is not a controlled unitary, for any $j_1$. Hence from \cite{cy13} and that $U$ has Schmidt rank $3$, $U_{j_1:\overline{j_1}}$ has Schmidt rank exactly $3$. Thus the space $\lin\{A_{i,j_1}\}_{i=1}^3$ has dimension three and is the $\cA_{j_1}$ space of $U$. Therefore from (i), this space is spanned by three unitary matrices. Assertion (ii.a) holds.

To prove (ii.b), suppose there are integers $j_1,\cdots,j_k$ such that $U_{j_1,\cdots,j_k:\overline{j_1,\cdots,j_k}}$ has Schmidt rank one or two. Up to the exchange of subscripts, we may assume that the three product operators $\{A_{i,j_1}\ox\cdots\ox A_{i,j_k}\}_{i=1,2,3}$ span a 1-dimensional or 2-dimensional space. Then $U$ is a controlled unitary, according to Lemma \ref{le:multiunitary} (iii). It is a contradiction with the condition $k\ge2$. Thus (ii.b) follows.

(iii) Since $k\ge3$ implies $k\ge2$, all assertions in (ii) apply to $U$.
Assume there are at most $k-1$ integers, say $l=1,\cdots,k-1$ such that for each $l$, the set $\{A_{i,l}\}_{i=1,2,3}$ contains a singular matrix. It implies that any matrix in the set $\{A_{i,l'}\}_{i=1,2,3}$ with $l'>k-1$ is invertible.
Since $k\ge3$, (ii.b) implies that the bipartite unitaries $U_{\overline{1,\cdots,k-2}:1,\cdots,k-2}$ and $U_{\overline{1,\cdots,k-1}:1,\cdots,k-1}$ have both Schmidt rank three. They are controlled unitaries by Theorem \ref{thm:sch3}. Since $\lc {n \over 2} \rc \ge k\ge3$, they are controlled by the systems $\cA_{k-1,\cdots,n}$ and $\cA_{k,\cdots,n}$, respectively. Let $C_i=A_{i,k-1}\ox B_i$, where $B_i=A_{i,k}\ox\cdots \ox A_{i,n}$ for $i=1,2,3$.
From Lemma \ref{le:controlunitary} (v), the operators $B_i B_j^\dg$ (resp. $C_i C_j^\dg$), $\forall i,j$ are all normal and commute with
each other, and the operators $B_i^\dg B_j$ (resp. $C_i^\dg C_j$), $\forall i, j$ are all normal
and commute with each other. We have
\bea
\label{eq:iiib1}
\big[
A_{i,k-1}A_{j,k-1}^\dg \ox B_i B_j^\dg,
A_{s,k-1}A_{t,k-1}^\dg \ox B_s B_t^\dg
\big] = 0,
\\
\label{eq:iiib2}
\big[
A_{i,k-1}^\dg A_{j,k-1} \ox B_i^\dg B_j,
A_{s,k-1}^\dg A_{t,k-1} \ox B_s^\dg B_t
\big] = 0,
\eea
for any $i,j,s,t\in[1,3]$. Since any matrix in the set $\{A_{i,l'}\}_{i=1,2,3}$ with $l'>k-1$ is invertible, the matrices $\{B_i\}_{i=1,2,3}$ are all invertible. It is known that the product of any two invertible matrices is invertible. Using these facts, \eqref{eq:iiib1} and \eqref{eq:iiib2}, we obtain that the operators $A_{i,k-1} A_{j,k-1}^\dg$, $\forall i,j$ are all normal and commute with
each other, and the operators $A_{i,k-1}^\dg A_{j,k-1}$, $\forall i, j$ are all normal
and commute with each other. So $\{A_{i,k-1}\}_{i=1,2,3}$ have simultaneous singular value decomposition. Then $U$ is a controlled unitary from $\cA_{k-1}$ side \cite{mm11}. It is a contradiction with $k\ge3$, so the assumption is wrong. The assertion (iii.a) follows.

Next we prove (iii.b). Evidently we have $\{i_l\}\sue\{1,2,3\}$. Suppose the two sets are equal. It means that there is a set of integers $S=\{l_1,l_2,l_3\}\in [1,n]$ such that $A_{1,l_1}$, $A_{2,l_2}$ and $A_{3,l_3}$ are all singular. From (iii.a), we can always find a set $S$ in which two integers are different. That is, $\abs{S}\ge2$.
If the inequality holds, let $\ket{b_i}$ be the states such that $\bra{b_i}A_{i,l_i}=0$ for $i=1,2,3$. Then $(\bra{b_1,b_2,b_3}\ox I_{\cA_{4,\cdots,n}})UU^\dg(\ket{b_1,b_2,b_3}\ox I_{\cA_{4,\cdots,n}})=0$. This is a contradiction with the fact that $U$ is unitary. So we have $\abs{S}=2$. Without loss of generality, we may assume $l_1\ne l_2=l_3$.
Because of $k\ge3$ and Theorem \ref{thm:sch3}, the $\cA_{l_1,l_2}$ space of $U$ is spanned by unitary matrices $U_1,U_2,U_3$ of Schmidt rank at most two. These matrices cannot have Schmidt rank one because of assertion (ii.a) and the fact that $A_{1,l_1}$, $A_{2,l_2}$ and $A_{3,l_3}=A_{3,l_2}$ are singular.
So $U_1,U_2,U_3$ all have Schmidt rank two.
Next, (ii) implies that any $U_i$ is the linear combination of two elements of $\{A_{i,l_1}\ox A_{i,l_2}\}_{i=1,2,3}$. Note that $A_{1,l_1}$, $A_{2,l_2}$ and $A_{3,l_3}=A_{3,l_2}$ are singular. From Lemma \ref{le:noninvertible} (i), any $U_i$ is the linear combination of $A_{2,l_1}\ox A_{2,l_2}$ and $A_{3,l_1}\ox A_{3,l_2}$. So the $U_1,U_2,U_3$ span a 2-dimensional space. It is a contradiction with the fact that they span the $\cA_{l_1,l_2}$ space of $U$, which has dimension three. Hence the two sets $\{i_l\}$ and $\{1,2,3\}$ are not equal. We obtain $\{i_l\}\su\{1,2,3\}$.

Next, suppose $\{i_l\}$ consists of exactly one integer in $\{1,2,3\}$, say $i_l=1$. It implies that any matrix $A_{i,l}$ with $i=2,3$ and $l=1,\cdots,n$ is invertible. Let
\bea
\label{eq:bj}
B_j=A_{j,3}\ox\cdots \ox A_{j,n}
\eea
for $j=1,2,3$. Assertion (ii.b) and Theorem \ref{thm:sch3} imply that $U_{12:\overline{12}}$ is a Schmidt-rank-three controlled unitary. It is controlled by the system $\cA_{3,\cdots,n}$, because of $k\ge3$. And since $U_{12:\overline{12}}=\sum^3_{j=1}A_{j,1}\ox A_{j,2} \ox B_j$ is a Schmidt decomposition across the bipartite cut $12:\overline{12}$, the matrices $\{B_j\}_{j=1,2,3}$ have simultaneous singular value decomposition. Note that $B_2,B_3$ are both invertible. So from Lemma~\ref{le:nonzerolinearcombi} (ii),
one of the two sets $\{A_{j,1}\}_{j=1,2,3}$ and $\{A_{j,2}\}_{j=1,2,3}$ have simultaneous singular value decomposition. We have $k=1$ and it is a contradiction with $k=3$. Thus, the set $\{i_l\}$ consists of exactly two distinct integers in $\{1,2,3\}$. The first assertion of (iii.b) follows.

Without loss of generality, let $\{i_l\}=\{1,2\}$. It implies that any matrix $A_{3,l}$ with $l=1,\cdots,n$ is invertible. By using (iii.a), we may assume that $A_{11}$ and $A_{22}$ are singular. Let $\ket{c_1},\ket{c_2}$ be the states such that $A_{11}\ket{c_1}=A_{22}\ket{c_2}=0$. Since $U$ is unitary, we have $(\bra{c_1,c_2}\ox I_{\cA_{3,\cdots,n}})U^\dg U(\ket{c_1,c_2}\ox I_{\cA_{3,\cdots,n}})=I_{\cA_{3,\cdots,n}}\propto A_{3,3}^\dg A_{3,3} \ox \cdots \ox A_{3,n}^\dg A_{3,n}$. So $A_{3,3},\cdots,A_{3,n}$ are all proportional to unitary matrices. Next, since $k\ge3$, the $\cA_{12}$ space of $U$ is spanned by three unitary matrices of Schmidt rank at most two. Lemma \ref{le:noninvertible} (i) implies that two of the three matrices are $b_{11} A_{11}\ox A_{12}+b_{12} A_{31}\ox A_{32}$ and $b_{21} A_{21}\ox A_{22}+b_{22} A_{31}\ox A_{32}$ with complex numbers $b_{11},b_{12},b_{21},b_{22}$, where $b_{12}b_{22}\ne0$, and at least one of $b_{11}$ and $b_{21}$ is nonzero. Since $A_{11}$ and $A_{22}$ are singular, Lemma \ref{le:noninvertible} (ii) implies that $A_{31}$ and $A_{32}$ are both proportional to unitary matrices. We have shown that any $A_{3l}$ with $l=1,\cdots,n$ is proportional to a unitary matrix. The second assertion of (iii.b) follows.

Lastly we prove (iii.c). We still use the definition of $B_j$ in \eqref{eq:bj}. An argument similar to those below \eqref{eq:bj} implies that $U=\sum^3_{j=1}A_{j,1}\ox A_{j,2} \ox B_j$ is a Schmidt-rank-three unitary, and the matrices $\{B_j\}_{j=1,2,3}$ have simultaneous singular value decomposition. We choose $x=3$ in the statement of (iii.b). Up to a local unitary on $\cH_{1,\cdots,n}$, we may assume $A_{31}=aI_{A_1}$ with some complex number $a$, $A_{32}=I_{A_2}$ and $B_3=I_{A_{3,\cdots,n}}$. Up to another unitary on $\cH_{\overline{12}}$, we may assume $B_1,B_2$ are both diagonal. Since $k\ge 3$, assertion (ii) implies that $U_{1:\overline{1}}$ is of Schmidt rank three, and it is controlled by the system $\cA_{\overline{1}}$. Lemma \ref{le:controlunitary} (v) implies
\bea
\label{eq:iiic}
\big[
A_{i,2}A_{j,2}^\dg \ox B_i B_j^\dg,
A_{s,2}A_{t,2}^\dg \ox B_s B_t^\dg
\big] = 0,
\eea
for any $i,j,s,t\in [1,3]$. Let $i=t=3$ and $j=s=1,2$. Then \eqref{eq:iiic} becomes $\big[
A_{j,2}^\dg \ox B_j^\dg,
A_{j,2} \ox B_j
\big] = 0$. Since both $B_1,B_2$ are diagonal, we have $[A_{j,2}^\dg,A_{j,2}]=0$. Thus $A_{j,2}$ is normal for $j=1,2,3$. By exchanging the system $\cA_2$ and any other $\cA_i$, one can similarly prove that $A_{j,i}$ is normal for $j=1,2,3$. Thus the assertion (iii.c) holds.
This completes the proof.
\epf

Assertion (iii.a) does not hold when $k=2$. A counterexample is the three-qubit $U^{(3)}$ in \eqref{eq:sr3threequbit}.
Now we are in a position to prove the main result of this section.

\bt
\label{thm:multi}
Every Schmidt-rank-three multipartite unitary is a controlled unitary, or is collectively controlled by two systems of $\cA_{1,\cdots,n}$.
\et
\bpf
Assume the assertion does not hold. Let $U$ be a Schmidt-rank-three $n$-partite unitary with $n\ge2$, and $U$ cannot be controlled by any two systems of $\cA_{1,\cdots,n}$. So all assertions in Lemma \ref{le:sr3=kcollective} apply to $U$. It follows from Theorem \ref{thm:sch3} that $n\ge5$. Using Lemma \ref{le:sr3=kcollective} (iii) and a suitable local unitary, we may assume
\bea
\label{eq:thmmulti}
U=A_1\ox A_2 \ox \cdots \ox A_n + D_1\ox D_2 \ox \cdots \ox D_n + a I_{\cA_1} \ox I_{\cA_2} \ox \cdots \ox I_{\cA_n},
\eea
where any $D_i$ is diagonal and the nonzero diagonal elements are in the upper left side of $D_i$, any $A_i$ is normal, and $a$ is a nonzero complex number. Since $U$ cannot be controlled by any two systems of $\cA_{1,\cdots,n}$, Lemma \ref{le:sr3=kcollective} implies that $U_{12:\overline{12}}$ is a Schmidt-rank-three bipartite unitary. It is a controlled unitary controlled by the system $\cA_{\overline{12}}$ by virtue of Theorem \ref{thm:sch3}. Thus the three product matrices $B_1=A_3\ox\cdots\ox A_n$, $B_2=D_3\ox\cdots\ox D_n$, and $B_3=I_{\cA_3} \ox \cdots \ox I_{\cA_n}$ have simultaneous singular value decomposition. Let $W$ and $V$ be unitaries on $\cH_{3,\cdots,n}$ such that $W B_i V =E_i$, $i=1,2,3$, where $E_i$ are all diagonal. In particular $E_3=W B_3 V=W V$ is a diagonal unitary. Hence $E_i E_3^\dg=W B_i V E_3^\dg$ are all diagonal. Note that $W B_3 V E_3^\dg=E_3 E_3^\dg =I$, and since $B_3=I$, we have $W^\dg = V E_3^\dg$. Thus $W B_i W^\dg $ are all diagonal. If $(W B_1 W^\dg)(W B_2 W^\dg)\ne0$, then there is a nonzero diagonal element in the same position of $W B_1 W^\dg,W B_2 W^\dg$ and $B_3$, respectively. Let this position be presented by the projector $\proj{l}$. Then the fact $U^\dg U=I$ implies $(I_{\cA_{1,2}}\ox\bra{l}) U^\dg U (I_{\cA_{1,2}}\ox \ket{l})=I_{\cA_{1,2}}$. So the $\cA_{12}$ space of $U$ contains a Schmidt-rank-three unitary, and it is a contradiction with $k\ge3$ and Theorem \ref{thm:sch3}. So we have $(W B_1 W^\dg)(W B_2 W^\dg)=0$, namely $B_1B_2=0$. It implies that there is an integer $i\in[3,n]$ such that $A_iD_i=0$. Since $A_i\ne0$, $D_i$ is singular. We may assume $D_i=D\op 0$ where $D$ is an $s\times s$ invertible diagonal matrix and $s<d_i$. The equation $A_iD_i=0$ implies that $A_i=\left(
                   \begin{array}{cc}
                     0 & C_1 \\
                     0 & C_2
                                       \end{array}
                 \right)$,
where $C_1,C_2$ are two blocks of size $s \times (d_i-s)$ and $(d_i-s) \times (d_i-s)$, respectively. Recall that $A_i$ is normal. We have $C_1=0$, and $C_2$ is also normal. Hence, the three matrices $A_i,D_i$ and $I_{\cA_i}$ are simultaneously diagonalizable. Eq. \eqref{eq:thmmulti} implies that $U$ is a controlled unitary. It is a contradiction with the assumption. This completes the proof.
\epf

The theorem implies that a Schmidt-rank-three multipartite controlled unitary can be characterized through the collaboration of two systems.
For example, we have seen that the Schmidt-rank-three 3-qubit unitary operation $U^{(3)}$ in \eqref{eq:sr3threequbit} can be collectively controlled by any two systems of the operation. Consider the Schmidt-rank-three 5-qubit unitary $V=I_2\ox I_2 \ox U^{(3)}$. Using Lemma \ref{le:controlunitary} (v), one can show that $V$ cannot be controlled by the system $\cA_{23}$. Hence in spite of Theorem \ref{thm:multi}, a Schmidt-rank-three multipartite unitary may be not collectively controlled by two \emph{random} systems.

Based on Theorem \ref{thm:multi} we obtain the following result. We shall refer to ``union'' as the case that a multipartite unitary is controlled by the union of a few systems, and may also be controlled by fewer systems. This is different from ``collectively control'' defined on in Sec. \ref{sec:intro}.
\bcr
\label{cr:multi}
Every Schmidt-rank-three $n$-partite unitary with $n\ge3$ is controlled by the union of two systems of $\cA_{1,\cdots,n}$. There exists a non-controlled Schmidt-rank-three tripartite unitary, i.e., it is collectively controlled by two systems of $\cA_{123}$.
\ecr
\bpf
The second assertion follows from \eqref{eq:sr3threequbit}. It suffices to prove the first assertion. Let $U$ be a Schmidt-rank-three $n$-partite unitary with $n\ge3$. Using Theorem \ref{thm:multi}, it suffices to prove the assertion when $U$ is a controlled unitary. Without loss of generality, we may assume that $U$ is controlled by system $\cA_1$. Lemma \ref{le:multiunitary} (ii) implies that $U$ is locally equivalent to $V=\sum^{3}_{i=1} A_{i,1} \ox A_{i,2}\ox\cdots\ox A_{i,n}$, where each $A_{i,1}$ is a diagonal matrix on $\cH_1$. By rewriting $V$ we have
\bea
\label{eq:vi}
V=\sum^{d_1}_{i=1}\proj{i}\ox V_i,
\eea
where each $V_i\in \lin\{A_{i,2}\ox\cdots\ox A_{i,n}\}_{i=1,2,3}$ is a unitary on the space $\cH_{\overline{1}}$. Since the $\{A_{i,2}\ox\cdots\ox A_{i,n}\}_{i=1,2,3}$ are linearly independent, each $V_i$ has Schmidt rank at most three. If some $V_j$ has Schmidt rank three, then (a)
$V_j$ is a controlled unitary controlled by one system $\cA_k$ of $\cA_{2,\cdots,n}$, or (b) $V_j$ has to be collectively controlled by two systems $\cA_{l,m}$ of $\cA_{2,\cdots,n}$ by Theorem \ref{thm:multi}. In case (a), note that $V_j\in \lin\{A_{i,2}\ox\cdots\ox A_{i,n}\}_{i=1,2,3}$ has Schmidt rank three. Lemma \ref{le:multiunitary} (ii) implies that the three matrices $A_{1,k},A_{2,k}$ and $A_{3,k}$ have simultaneous singular value decomposition. So $V$ is collectively controlled by the system $\cA_{1,k}$, and the assertion follows. In case (b), Lemma \ref{le:sr3=kcollective} (ii) implies that $V_j$ is a Schmidt-rank-three bipartite unitary on any bipartite cut. So the three product matrices $A_{i,l}\ox A_{i,m}$, $i=1,2,3$ have simultaneous singular value decomposition. So $V$ is collectively controlled by the system $\cA_{l,m}$, and the assertion follows.

It remains to prove the assertion when each $V_i$ has Schmidt rank at most two. Ref. \cite{cy13} implies that each $V_i$ is a controlled unitary controlled by any system in $\cA_{2,\cdots,n}$. Let $W_i,X_i$ be two unitary matrices on $\cH_2$ such that the $\cA_2$ space of $(W_i\ox I_{\cA_{3,\cdots,n}})V_i(X_i\ox I_{\cA_{3,\cdots,n}})$ is spanned by a diagonal basis. Let $W=\sum^{d_1}_{i=1}\proj{i}\ox W_i$ and $X=\sum^{d_1}_{i=1}\proj{i}\ox X_i$ be two unitary operators on $\cH_{12}$. From \eqref{eq:vi}, the $\cA_{12}$ space of $(W\ox I_{\cA_{3,\cdots,n}})V(X\ox I_{\cA_{3,\cdots,n}})$ is spanned by a diagonal basis. So $V$ is collectively controlled by $\cA_{12}$. This completes the proof.
\epf

To summarize, if a Schmidt-rank-three multipartite unitary $U$ is a controlled unitary by a system say $\cA_1$, then it is also controlled by the union of two systems. Nevertheless, we do not know whether the two systems always contain $\cA_1$.

On the other hand, if $U$ is collectively controlled by two systems, then it may be not collectively controlled by any one or any three systems. An example is when $U=U^{(n)}$ with any odd $n\ge5$, see
\eqref{eq:sr3nqubit}. On the other hand, $U^{(n)}$ is collectively controlled by any two systems of $\cA_{1,\cdots,n}$ by its symmetry and Theorem \ref{thm:multi}.

\subsection{Schmidt-rank-three multiqubit unitary operators}
\label{subsect:multiqubit}

Multiqubit controlled gates can be decomposed into certain elementary gates \cite{bbc95}, and are more useful than the general multipartite unitaries. For example,
controlled NOT (CNOT) gates are essential for the construction of universal quantum two-qubit gates used in quantum computing both theoretically and experimentally \cite{bbc95,sw95,cz00}. Recently CNOT gates have been proved to be implemented by trapped ions controlled by fully
overlapping laser pulses \cite{msv14}. Multiqubit graph and cluster states
for one-way quantum computing are generated by
controlled-Z gates \cite{br01}. Controlled phase gates have also
been used to construct the mutually unbiased bases (MUBs)
\cite{wpz11}. Motivated by these applications and results in previous sections, we explore the Schmidt-rank-three multiqubit unitaries.

It is known that a two-qubit unitary cannot have Schmidt rank three \cite{Nielsen03}. In \eqref{eq:sr3threequbit}, we have presented the Schmidt-rank-three non-controlled three-qubit unitary constructed in \cite{cy13}. Indeed, we can show that the following Schmidt-rank-three $n$-qubit unitary
\bea
\label{eq:sr3nqubit}
U^{(n)}={1\over\sqrt3} [(\s_0)^{\ox n} + i(\s_1)^{\ox n} + i(\s_3)^{\ox n} ]
\eea
with any odd $n\ge3$ is not a controlled unitary. On the other hand, constructing a non-controlled Schmidt-rank-three 4-qubit unitary turns out to be impossible. For any even $n\ge4$, we will show that the Schmidt-rank-three $n$-qubit unitary is a controlled unitary, see Proposition \ref{pp:evenqubit}. This is the main result of this section. For this purpose we present the following preliminary lemma.

\bl
\label{le:44}
Let $V=\left(
                   \begin{array}{cccc}
                     0 & 0 & 0 & 0 \\
                     0 & a & b & c \\
                     0 & d & e & f \\
                     0 & g & h & i \\
                                       \end{array}
                 \right)$ be a two-qubit operator.
If $V$ has rank one and Schmidt rank one, then one of the following four conditions is satisfied:
\\
(i) $a=b=c=e=h=0$, $di=fg$ and $d\ne0$.
\\
(ii) $a=d=e=f=g=0$, $bi=ch$ and $b\ne0$.
\\
(iii) $a=b=c=d=g=0$, $ei=fh$, and one of $e,f,h$ is nonzero.
\\
(iv) $b=d=e=f=h=0$ and $ai=cg$.
\el
\bpf
The proof is by straightforward computation. We investigate three cases: $d\ne0$, $b\ne0$ and $d=b=0$. Since $V$ has rank one and Schmidt rank one, the first and second cases imply (i) and (ii), respectively. In the third case, we study two subcases: one of $e,f,h$ is nonzero, and they are all zero. They imply (iii) and (iv), respectively.
This completes the proof.
\epf

\bpp
\label{pp:evenqubit}
For any even integer $m\ge4$, the Schmidt-rank-three $m$-qubit unitary is a controlled unitary.
\epp
\bpf
Suppose the assertion is not true. It implies three items as the hypotheses. First, let $n$ be the minimum even integer, such that $n\ge4$ and
\bea
\label{eq:u}
U=\sum^3_{i=1}A_{i,1}\ox A_{i,2}\ox A_{i,3}\ox \cdots \ox A_{i,n}
\eea
is a non-controlled Schmidt-rank-three $n$-qubit unitary.
So Lemma \ref{le:sr3=kcollective} (ii) applies to $U$. It implies the second hypothesis: the three $2\times2$ matrices $A_{1,j},A_{2,j},A_{3,j}$ with any $j\in [1,n]$ are linearly independent, and do not have simultaneous singular value decomposition. Third, $U$ is a Schmidt-rank-three bipartite unitary on any bipartite cut. These hypotheses do not change if we switch any two systems of $\cA_{123\cdots n}$, or if we switch any two product operators $A_{i,1}\ox A_{i,2}\ox A_{i,3}\ox \cdots \ox A_{i,n}$ and $A_{j,1}\ox A_{j,2}\ox A_{j,3}\ox \cdots \ox A_{j,n}$ by relabeling the subscripts $i,j$. Since $U$ is not a controlled unitary, Theorem \ref{thm:multi} implies that $U$ is collectively controlled by two systems of $\cA_{123\cdots n}$. We can switch the systems so that $U$ is controlled by the system $\cA_{12}$. The third hypothesis implies that the three matrices in the set $S=\{A_{i,1}\ox A_{i,2}\}_{i=1,2,3}$ have simultaneous singular value decomposition.

We discuss in this paragraph three cases in terms of the ranks of the matrices in $S$, and will discuss the remaining fourth case in the next paragraph. First, if all matrices in $S$ have rank one, then $UU^\dg\ne I$. It is a contradiction with the first hypothesis that $U$ is unitary. So one matrix in  $S$ has rank at least two. Next, if two matrices in $S$ are invertible (with rank four), then \eqref{eq:u} implies that there are two states $\ket{a},\ket{b}\in\cH_{12}$, and nonzero complex numbers $c_1,c_2,c_3$, such that $(\bra{a}\ox I_{\cA_{\overline{12}}})U=\bra{b} \ox U'$ where $U'=\sum^3_{i=1}c_i A_{i,3}\ox \cdots \ox A_{i,n}$. Since $U$ is unitary, $U'$ is a $(n-2)$-qubit unitary. The third hypothesis implies that $U'$ has Schmidt rank three by Lemma \ref{le:multiunitary} (i). Since $U$ is not a controlled unitary, neither is $U'$. Since $n\ge4$, we obtain a contradiction with the first hypothesis that $n$ is minimum. Thus two matrices in $S$ are singular.
Third, if there is a matrix of rank two in $S$, then up to the relabeling of subscripts and the switch of systems $\cA_1$ and $\cA_2$, we may assume such matrix is $A_{1,1}\ox A_{1,2}$, and it satisfies $\rank A_{1,1}=1$ and $\rank A_{1,2}=2$.
Using a suitable product unitary on $\cH_{12}$ and absorbing a suitable factor into $A_{1,3}$, we may assume $A_{1,1}=\diag(1,0)$ and $A_{1,2}=\diag(1,c)$ with some positive number $c$. Note that the matrices in $S$ have simultaneous singular value decomposition. Let $W,X$ be two unitary matrices on $\cH_{12}$ such that $W(A_{i,1}\ox A_{i,2})X$ is diagonal for $i=1,2,3$. In particular, we may choose $W,X$ such that $W(A_{1,1}\ox A_{1,2})X=A_{1,1}\ox A_{1,2}$.
Since $A_{1,1}\ox A_{1,2}=\diag(1,c,0,0)$ and $c>0$, Lemma \ref{le:normal} (iii) implies that $W$ and $X$ are both the direct sum of two $2\times2$ unitary matrices. Using this fact and noting that $W(A_{i,1}\ox A_{i,2})X$ is diagonal for $i=2,3$, we obtain that $A_{i,1}\ox A_{i,2}$ is the direct sum of two $2\times2$ matrices. Thus $A_{2,1}$ and $A_{3,1}$ are both diagonal. Since $A_{1,1}=\diag(1,0)$, $U$ is a controlled unitary. This is a contradiction with the first hypothesis. So any matrix in $S$ does not have rank two.

The three cases in the last paragraph imply that the three matrices in $S$ have to respectively have rank one, one and four. Up to a relabeling of subscripts, we may assume
\bea
\label{eq:rank}
&&
\rank A_{1,j}= \rank A_{2,j}= 1,
\notag\\
&&
\rank A_{3,j}=2,
\eea
with $j=1,2$. Up to a product unitary on $\cH_{12}$, we may assume
\bea
\label{eq:a11a12}
A_{1,1}= A_{1,2}=\diag(1,0).
\eea
Note that $U$ is controlled by $\cA_{12}$. Let $W_1,X_1$ be two unitaries on $\cH_{12}$ such that
\bea
\label{eq:wa11a12x}
W_1(A_{1,1}\ox A_{1,2})X_1=A_{1,1}\ox A_{1,2},
\eea
and $W_1(A_{i,1}\ox A_{i,2})X_1$ with $i=2,3$ are still diagonal.
Eqs. \eqref{eq:a11a12}, \eqref{eq:wa11a12x}, and Lemma \ref{le:normal} (iii) imply that $W_1$ and $X_1$ are both the direct sum of integer $1$ and a $3\times3$ unitary matrices.
Since $W_1(A_{3,1}\ox A_{3,2})X_1$ is diagonal, $A_{3,1}\ox A_{3,2}$ is the direct sum of a complex number and a $3
\times3$ matrix $B_1$. Since $A_{3,1}\ox A_{3,2}$ is invertible, $B_1$ is diagonal. So $A_{3,1}$ and $A_{3,2}$ are both diagonal. Up to a diagonal product unitary on $\cH_{12}$, we may assume that $A_{3,1}$ and $A_{3,2}$ are both positive definite without changing $A_{1,1}$ and $A_{1,2}$ in \eqref{eq:a11a12}. Note that $U$ is still a controlled unitary controlled by $\cA_{12}$. From \eqref{eq:a11a12} and Lemma \ref{le:normal} (iii), there exist two unitaries
\bea
\label{eq:w'x'}
W' = 1 \op W_2,
~~~
X' = 1 \op X_2
\eea
on $\cH_{12}$ such that
\bea
\label{eq:w'a11a12x'1}
W'(A_{1,1}\ox A_{1,2})X' &=& A_{1,1}\ox A_{1,2},
\notag
\\
\label{eq:w'a11a12x'2}
W'(A_{3,1}\ox A_{3,2})X' &=& A_{3,1}\ox A_{3,2},
\eea
and $W'(A_{2,1}\ox A_{2,2})X'$ is diagonal. Thus $A_{2,1}\ox A_{2,2}$ is the direct sum of a complex number $x$ and a $3\times3$ matrix $B_2$. The second hypothesis implies that $A_{2,1}\ox A_{2,2}$ is not parallel to $A_{1,1}\ox A_{1,2}$. These facts, \eqref{eq:rank} and \eqref{eq:w'x'} imply $x=0$.
Thus $A_{2,1}\ox A_{2,2}$ becomes the matrix $V$ in Lemma \ref{le:44}.  It satisfies one of the four conditions (i-iv) in Lemma \ref{le:44}.
From \eqref{eq:a11a12} and the paragraph below \eqref{eq:wa11a12x}, the four matrices $A_{1,1},A_{3,1},A_{1,2}$ and $A_{3,2}$ are all diagonal.
Since $U$ is not a controlled unitary, conditions (iii) and (iv) in Lemma \ref{le:44} are excluded.
Next, either condition (i) or (ii) in Lemma \ref{le:44} implies that $A_{2,1}\ox A_{2,2}$ is not normal. Since $W'(A_{2,1}\ox A_{2,2})X'$ is diagonal, we have $W'\ne(X')^\dg$. Since
$A_{3,1}\ox A_{3,2}$ is diagonal positive definite, we have a contradiction because of \eqref{eq:w'a11a12x'2} and Lemma \ref{le:normal} (iv).

We have excluded all possible cases of the matrices in $S$. So the hypothesis is wrong, and the assertion holds. This completes the proof.
\epf

We claim that a nontrivial Schmidt-rank-three $n$-qubit unitary exists for every $n\ge3$, where ``nontrivial'' means not the tensor product of a one-qubit unitary and a Schmidt-rank-three $(n-1)$-qubit unitary. If $n$ is odd then the claim follows from \eqref{eq:sr3nqubit}. For even $n$, let $V^{(n-1)}=(U^{(n-1)})^\dg$ where the latter is defined in \eqref{eq:sr3nqubit}. Then $\proj{1}\ox U^{(n-1)} +\proj{2}\ox  V^{(n-1)}$
is a Schmidt-rank-three $n$-qubit unitary. So the claim follows.

Proposition \ref{pp:evenqubit} implies that the Schmidt-rank-three $n$-qubit unitaries with odd and even $n$ have different control properties. The reason that makes the difference might be from the mathematical structure of multiqubit unitaries, but a decisive proof is not known yet.

\subsection{Entanglement cost of implementing a bipartite unitary}
\label{sec:entanglementcost}

The following lemma generalizes \cite[Lemma 9]{cy14}. It follows simply from a few protocols of implementation and Theorem \ref{thm:sch3}, as explained after the lemma.

 \bl
 \label{le:costSR3}
Let $d_A\le d_B$. Any bipartite unitary of Schmidt rank three can be
implemented by using LOCC and the maximally entangled state
$\ket{\Ps_{k}}$, where $k=\min\{d_A^2,d_B\}$.
 \el

The $d_A^2$ term is from using teleportation \cite{bbc93} twice:
Alice teleports her input system to Bob, and Bob does the unitary locally, and teleports back the part of the output system belonging to Alice to her.
This requires two maximally entangled states $\ket{\Ps_{d_A}}$ ($d_A\le d_B)$, which contains $2\log_2 d_A$ ebits \cite{ygc10}.

The $d_B$ term is from the protocol for controlled unitaries in ref.~\cite{ygc10}, which uses a maximally entangled state of Schmidt rank equal to the number of terms in the expression of the unitary in the controlled form.

From this lemma, $\log_2 d_B$ ebits is an upper bound of the
amount of entanglement needed to implement a bipartite
unitary of Schmidt rank three.  As discussed in \cite{cy14}, this upper bound can be saturated for some unitary with $d_A=2$, $d_B=3$.

It is still an open question whether there is a Schmidt-rank-three unitary that needs more than $\log_2 3$ ebits to exactly implement using LOCC.  This is a question about the lower bound of entanglement cost of unitaries, and some known results are in Soeda \textit{et al} \cite{stm11} and Stahlke \textit{et al} \cite{sg11}.  These results suggest the interesting case to look at is when the resource state has Schmidt rank greater than that of the unitary, but still with relatively small entanglement.

On the probabilistic implementation of unitaries of small Schmidt rank, the protocol involving gate-teleportation that implements some types of two-qubit or two-qudit unitaries in \cite{mr13} can be generalized to a protocol that probabilistically implements an arbitrary unitary acting on a $(d_A\times d_B)$-dimensional space, by using the generalized Bell-state measurements and removing the final corrections. This would implement any Schmidt-rank-$r$ unitary with probability $1/(d_A d_B)^2$ using a maximally entangled state of Schmidt rank $r$, but in the cases of failure, it is hard to recover the desired unitary via local corrections.

\subsection{Connection with the Schmidt number}
\label{sec:schmidtnumber}

The Schmidt number of a bipartite quantum state is firstly introduced in \cite{th00}. It becomes the standard Schmidt rank when the state is pure. The Schmidt number is an entanglement monotone under LOCC.
Recently, a device independent Schmidt rank witness for bipartite pure states has been proposed by using the Hardy paradox \cite{mrb14}. So the Schmidt rank is to some degree an observable quantity, and it is well known that it has various applications in quantum information. Similar statements can be said for the Schmidt number. In this subsection we explore the relation between the Schmidt number and controlled unitaries. For the convenience of readers, we review the definition from \cite{sb00}.
\bd\label{def:schnumber}
Given the density matrix $\rho$ of a bipartite system and all its possible decompositions in terms of pure states, namely
$\rho=\sum_i p_i \ketbra{\psi_i^{r_i}}{\psi_i^{r_i}}$, where $r_i$ denotes the Schmidt rank of $\ket{\psi_i^{r_i}}$, the Schmidt number of $\rho$ is
defined as $\min\{\max_i\{r_i\}\}$ where the minimum is taken over all decompositions.
\ed

For example, the separable state has Schmidt number one in terms of the definition. Since projectors are extremal in the set of normalized quantum states (trace-one positive semidefinite matrices), the Schmidt number of a pure state is equal to its Schmidt rank. Below we study how the Schmidt number of the output state of a bipartite unitary is related to the Schmidt rank of the unitary, when the input state is restricted to be a separable state. We have not obtained an exact relationship, but some partial results are presented below.

\textbf{Observation.} For any $r\ge 1$, all diagonal bipartite unitaries of Schmidt rank $r$ satisfy that there is a pure product input state such that the corresponding output state has Schmidt rank $r$. The input state can be chosen to be $\ket{+}_A\ket{+}_B$, where $\ket{+}_A:=\frac{1}{\sqrt{d_A}}\sum_{j=1}^{d_A} \ket{j}$, and $\ket{+}_B:=\frac{1}{\sqrt{d_B}}\sum_{j=1}^{d_B} \ket{j}$.

\bl
\label{le:schnum}
Suppose a bipartite unitary is of Schmidt rank $r$, and it acts on a separable input state.\\
(i) The Schmidt number of the output state is equal to or smaller than $r$.\\
(ii) The equality is always achievable by some suitable input state when ancillas of sufficient size are allowed as part of the extended input. In particular, when $r\le 3$, the controlling party does not need any ancilla.\\
(iii) When ancillas are not allowed, the equality is achieved by some suitable input state (dependent on the unitary) for any bipartite unitary of Schmidt rank $r\le 2$, and not achieved for some bipartite unitary of Schmidt rank $r$ where $r$ is any given integer greater than $2$.
\el
\bpf
(i) Let $U=\sum^r_{j=1}A_j\ox B_j$ be a Schmidt-rank-$r$ bipartite unitary. Suppose the separable input state is $\rho=\sum_i \ketbra{\a_i}{\a_i}\ox\ketbra{\b_i}{\b_i}$.
For each pure-state component $\ket{\a_i}\ox\ket{\b_i}$, the corresponding output is $\sum_{j=1}^r A_j\ket{\a_i}\ox B_j\ket{\b_i}$, so it is of Schmidt rank not greater than $r$. By the definition of the Schmidt number, the overall output state has Schmidt number not greater than $r$. This conclusion holds regardless of whether ancillas are allowed for the input state, since the tensor product of $U$ with identity operators on the ancillas still has Schmidt rank $r$. This proves (i).

(ii) If we are allowed to add ancillas $\bar A$ and $\bar B$ with sizes equal to that of $A$ and $B$, respectively, then there is an input state $\frac{1}{\sqrt{d_A d_B}}\sum_{j=1}^{d_A} \ket{j}_A \ket{j}_{\bar A} \ox \sum_{k=1}^{d_B} \ket{k}_B \ket{k}_{\bar B}$, such that the output state on $A\bar A B\bar B$ has Schmidt rank $r$ across the $A\bar A : B\bar B$ cut, equal to the Schmidt rank of $U$. This proves the first part of (ii). For the second part, from Theorem \ref{thm:sch3} and \cite{cy13}, the unitary is a controlled unitary, so up to local unitaries and a possible swap of the $A$,$B$ systems, the unitary is of the form $U=\sum^n_{j=1} \ketbra{j}{j}\ox V_j$, where $n\ge r$, and $V_j$ ($1\le j\le n$) are unitaries, with $r$ of them being linearly independent. Choose the input state to be $\frac{1}{\sqrt{d_A d_B}}\sum_{j=1}^{d_A} \ket{j}_A \ox \sum_{k=1}^{d_B} \ket{k}_B \ket{k}_{\bar B}$, then the output state on $A B\bar B$ has Schmidt rank $r$ across the $A : B\bar B$ cut, equal to the Schmidt rank of $U$. The effect of the local unitaries is to change the form of the input state above, but does not affect the size of the possible ancillas. This proves the second part of (ii).

(iii) In the following we assume ancillas are not allowed. The equality can obviously be achieved when $r=1$. From \cite{cy13}, all bipartite unitaries of Schmidt rank $2$ is locally equivalent to a diagonal unitary, thus from the Observation above, there is a product pure input state with the output state having Schmidt rank $2$. In the case $r=3$, $U$ may be a controlled unitary on a $d_A\times 2$ system, then the maximum Schmidt number of the output state is $2$, which is less than the Schmidt rank of the unitary. In the case $r=4$, let $U$ be the SWAP gate on two qubits, then the Schmidt number of the output state is always $1$ for any separable input state. For $r>4$, similar examples of unitaries with $\min\{d_A,d_B\}<r$ can be constructed so that the maximum Schmidt number of the output is not greater than $\min\{d_A,d_B\}$ and thus less than $r$, when the input is separable. This completes the proof.
\epf

\smallskip
From the proof above, we see that for the question of whether the equality is achievable, the ancillas become important since addition of ancillas increase the dimensions of the local Hilbert spaces which are upper bounds for the Schmidt number of the output state.

Define the Schmidt rank of a multipartite pure state to be the minimum number of pure product states that sum to the given state. By defining the Schmidt number of a multipartite mixed state based on it, similar to Definition~\ref{def:schnumber}, we can consider the generalizations of the results above to the multipartite case. The ``Observation'' and Lemma~\ref{le:schnum} (i) can be straightforwardly generalized to the multipartite case, but Lemma~\ref{le:schnum} (ii)(iii) may need some modifications in generalization, and we leave it for future study.

\subsection{Bipartite unitaries studied in Ref. \cite{cy13}}
\label{sec:cy13}

The Schmidt-rank-two multipartite unitary operators have been thoroughly studied in \cite{cy13}. Apart from the usefulness mentioned before, they are also useful for other quantum technology. For example, the unitary $\proj{0}\ox (I_2)^{\ox n}+\proj{1}\ox (\s_1)^{\ox n}$ can amplify a single spin by set of ancillary spins  \cite[Eq. (1)]{cfb11}. So it is meaningful to understand the Schmidt-rank-two unitaries from another point of view. In this subsection, we give an alternative proof of one of the main results in \cite{cy13}, using only the linear algebra in Sec. \ref{sec:pre}.
\bl
\label{le:sch2}
Any bipartite unitary of Schmidt rank two is a controlled unitary controlled from either side, and is a diagonal unitary up to local unitaries.
\el
\bpf
Let $U=\sum^2_{j=1}A_j\ox B_j$ be a Schmidt-rank-2 bipartite unitary. From Lemma \ref{le:nonsingular} (i) we may assume that $A_1$ is singular. Up to local unitaries on $\cH_A$, we may assume $A_1\ket{1}=0$. Since $(\bra{1}\ox I_B)U^\dg U(\ket{1}\ox I_B)=I_B$, we have
\bea\label{eq:sch21}
I_B=x B_2^\dg B_2,
\eea
where $x$ is a positive real number. Hence $B_2$ is proportional to a unitary.
There is a complex number $z$ such that the linear combination $A_2'=z A_1+A_2$ satisfies $\tr(A_2'^\dg A_1)=0$, hence $U=A_1\ox B_1'+A_2'\ox B_2$, where $B_1'=B_1-z B_2$. The operators $B_1'$ and $B_2$ are linearly independent. By considering $U^\dg U=I$ and taking the partial trace over the $A$ side, we have
\bea\label{eq:sch22}
I_B=s B_1'^\dg B_1' + t B_2^\dg B_2,
\eea
where $s>0$, $t>0$. Hence from Eq.~\eqref{eq:sch21}, we have that $B_1'$ is proportional to a unitary. By absorbing factors into the A-side operators $A_1$ and $A_2'$, we have that both $B_1'$ and $B_2$ are unitaries. Up to local unitaries on $H_B$, we may assume $B_1'=I$ and $B_2$ is a diagonal unitary. Both operators are diagonal, thus $U$ is controlled from the $B$ side.

The argument above also works with the $A$ and $B$ sides swapped. Hence $U$ is a controlled unitary, controlled from either side. Up to local unitaries, $U=\sum_{j=1}^{d_A} \ketbra{j}{j}\ox C_j$, where $C_j$ are unitaries on $\cH_B$. Then we have $C_j\in\lin\{C_g,C_h\}$ for two distinct integers $g$, $h$ in $\{1,2,\cdots,d_A\}$, and up to local unitaries on $\cH_B$, we have that $C_g=I_B$, and $C_h$ is a diagonal unitary, hence all $C_j$ are diagonal, therefore $U$ is a diagonal unitary up to local unitaries. This completes the proof.
\epf

\bigskip

\section{\label{sec:conclusion} Conclusions}

We have shown that any bipartite unitary operator of Schmidt rank three
is locally equivalent to a controlled unitary. We have shown that LOCC and the $r\times r$ maximally entangled state of
$r=\min\{d_A^2,d_B\}$ (under the assumption that $d_A\le d_B$) are sufficient to implement such operators. We further show that any multipartite unitary operator of Schmidt rank three is a controlled unitary, or collectively controlled by two systems, regardless of the number of systems of this operator. We further show that the Schmidt-rank-three $n$-qubit unitary with any even $n\ge4$ is a controlled unitary. We also have found a connection between the Schmidt number and controlled unitaries, in terms of the separable inputs and ancillas. Using the methods in this paper we have retrieved a main result of \cite{cy13} in a different way.

There are many interesting open problems arising in this paper. It is expected that the technique and results developed here could be useful for characterizing nonlocal unitaries of larger Schmidt rank. Next, we have sort of characterized the Schmidt-rank-three multiqubit unitary in terms of $U^{(n)}$ and Proposition \ref{pp:evenqubit}. It is unknown whether multiqubit unitaries of higher Schmidt rank can be similarly characterized. This is related to an unproved intuition that the Schmidt-rank-three multiqubit unitary might have a simpler structure than the general Schmidt-rank-three multipartite unitary.
Another interesting problem is to construct a deeper connection between the nonlocal unitaries and Schmidt number of a bipartite or multipartite mixed state, extending the results in Sec.~\ref{sec:schmidtnumber}. We also do not know the entanglement cost of implementing a Schmidt-rank-three multipartite unitary operation. Finally, although we have shown that any Schmidt-rank-three multipartite unitary is collectively controlled by at most two systems, an efficient method for finding out the two systems is still lacking.

\bigskip

\section*{Acknowledgments}

We thank Scott Cohen, Joseph Fitzsimons and Yingkai Ouyang for useful discussions or helpful comments. This material is based on research funded by the Singapore National Research Foundation under NRF Grant No. NRF-NRFF2013-01.

\bibliographystyle{unsrt}

\bibliography{channelcontrol}

\end{document}